\documentclass[12 pt]{article}
\usepackage{cite}
\usepackage[dvips]{epsfig}
\usepackage{graphicx}
\graphicspath{{bilder/}}
\def\bare{\mathaccent28695}
\begin{document}
\title{Field Theory of Critical Behaviour in Driven Diffusive
Systems with Quenched Disorder}
\author{V. Becker and H.K. Janssen\\
Institut f\"ur Theoretische Physik III\\
Heinrich-Heine-Universit\"at D\"usseldorf\\
D-40225 D\"usseldorf, Germany}
\date{March 22, 1998}
\maketitle
We present a field theoretic renormalization group study for the
critical behaviour of a uniformly driven diffusive system with
quenched disorder, which is modelled by different kinds of potential
barriers between sites. Due to their symmetry properties, these
different realizations of the random potential barriers lead to three
different models for the phase transition to transverse order and to
one model for the phase transition to longitudinal order all belonging
to distinct universality classes. In these four models that have
different upper critical dimensions $ d_{c} $ we find the critical
scaling behaviour of the vertex functions in spatial dimensions $ d
< d_{c} $. Its deviation from purely diffusive behaviour is
characterized by the anomaly--exponent $ \eta $ that we calculate at
first and second order, respectively in $ \epsilon = d_{c} -d $. In
each model $ \eta $ turns out to be positive which means
superdiffusive spread of density fluctuations in the driving force direction.

\section{Introduction}
  
\begin{figure}[!ht]
    \begin{center}
      \leavevmode
      \epsfig{file=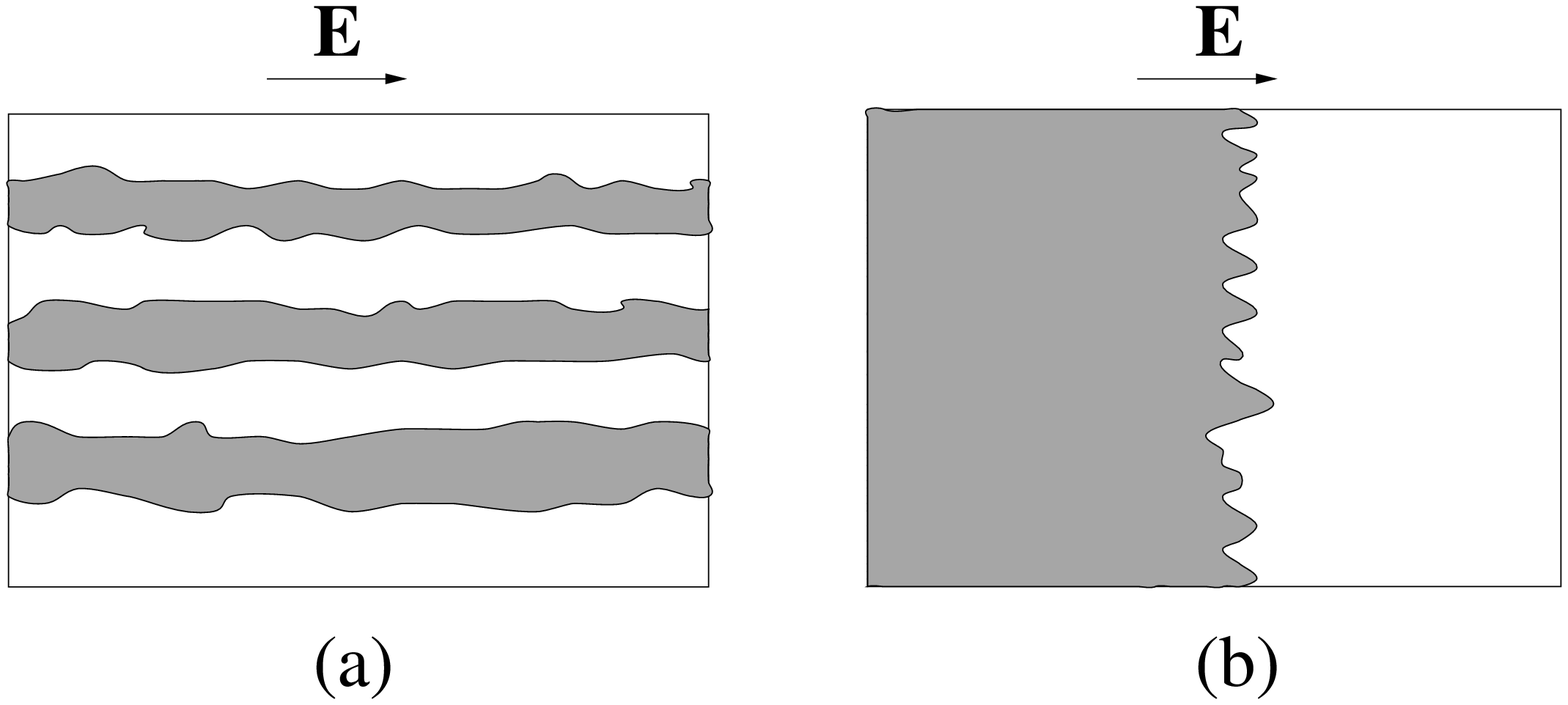, width=11cm}
       \caption{Typical configurations of transverse (a) and longitudinal (b)
      ordered phase. Regions of high density are shaded.}
     \label{uebergang}
    \end{center}
  \end{figure}

For more than a decade the long-time and critical behaviour of diffusive
systems subjected to a driving force has attracted considerable
interest. This is mainly caused by the richness of their highly
nontrivial features which generically result from the fact that, due to
the driving force, in general even the steady states of these systems
are far from thermal equilibrium, especially in the case of very
strong driving forces. Further, driven diffusive systems might be suitable
models for fast ionic conductors which was first suggested by Katz,
Lebowitz, and Spohn \cite{KLS1,KLS2}. A review of the different
investigations on driven diffusive systems and their
relations to other non--equilibrium systems is given by Schmittmann
and Zia \cite{SZ}.\\
We are interested in the diffusive motion of uniformly driven
particles with short range attractive interactions and hardcore
repulsion. With such an interaction driven diffusive systems show two
kinds of phase transitions. At the transverse (with respect to the
driving force $ {\bf E} $) phase transition the system changes from a
disordered state to an ordered one where the system is ordered in the
transverse direction and remains disordered in the longitudinal
(parallel to the driving force) direction. The typical configurations
are strips of high-- and low--density phase, arranged parallel to the
driving force ${\bf E}$ (Fig. \ref{uebergang}(a)). At the longitudinal phase
transition the systems changes from a disordered state to a state
where the system is ordered in the longitudinal direction and remains
disordered in the transverse directions. Here the typical
configurations are domains ("pancakes") of different phase with
interfaces perpendicular to the driving force (Fig. \ref{uebergang}(b)). 
In preceding papers the long--time and critical behaviour
of driven diffusion in an ordered medium \cite{JS1,JS2,LC} and the long--time
behaviour of driven diffusion in a medium with quenched disorder
\cite{BJ1} have been investigated by renormalized field theory. In the
present paper we study the effect of quenched disorder on the two
kinds of phase transitions in driven diffusive systems. Quenched
disorder is important in real systems, as it models impurities and defects.
We stress that this work completes the investigations of a
whole model class that is graphically shown in the summary
(Fig. \ref{ahnen1}). Note that the effects of quenched disorder in
randomly driven diffusive systems were recently analyzed
\cite{SchmBass,SchmLab}. There the 
driving force is itself a locally random variable with
respect to its amplitude and sign, whereas here the driving force is uniform.\\
The paper is organized as follows: In Section II we set up a Langevin
description of our system, directly on a mesoscopic length scale using
conservation laws and symmetry arguments. The quenched disorder is
modelled by random potential barriers between sites, and the
symmetry properties of the random potential play a crucial
role. Different realizations of the random potential are possible and
are argued to lead to different kinds of noise.
In Section III we study the transverse phase transition in the
convenient formalism of the dynamic functional. In Section IV the
longitudinal phase transition is analyzed. Section V contains the main
results and a graphical overview of the entire model class.

\section{Model Building}
\begin{figure}[!ht]
    \begin{center}
      \leavevmode
      \epsfig{file=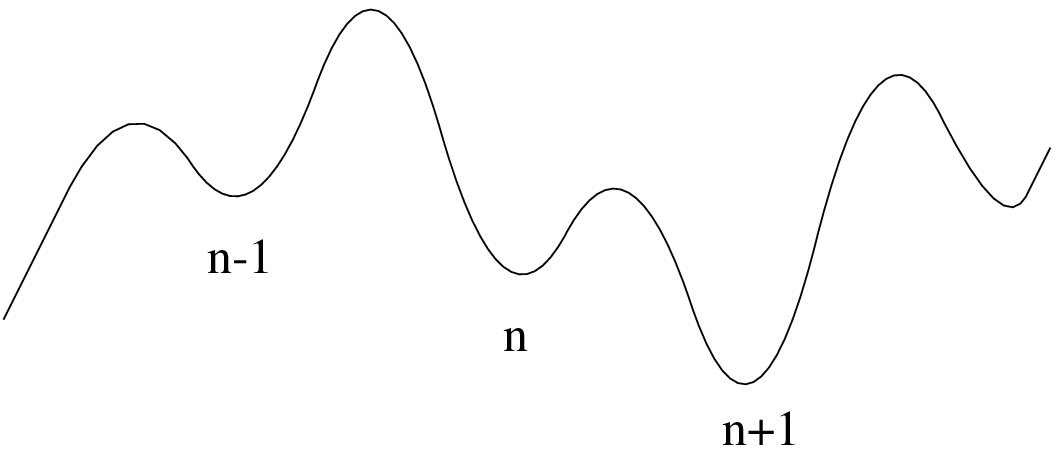, width=8cm}\\
(a)\\[10 mm]
      \epsfig{file=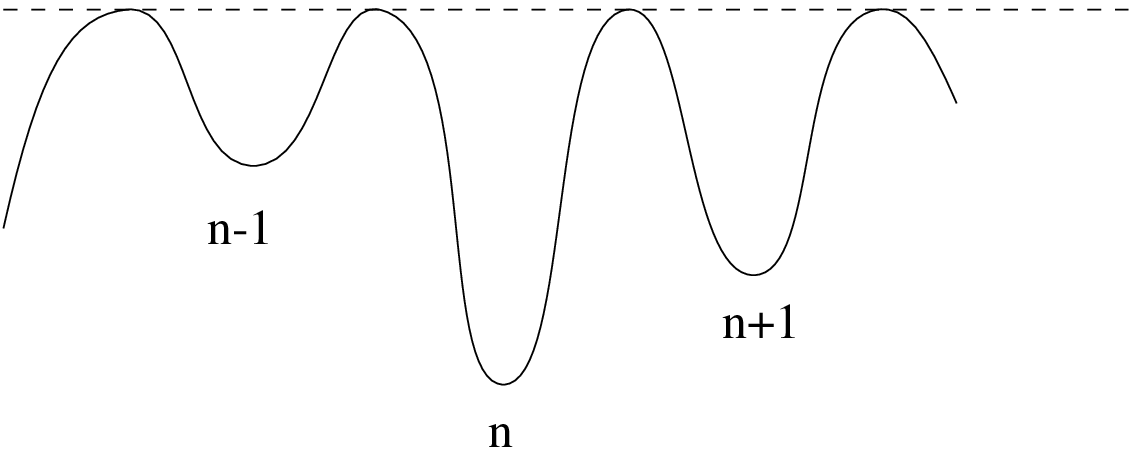, width=8cm}\\
(b)\\[10 mm]
      \epsfig{file=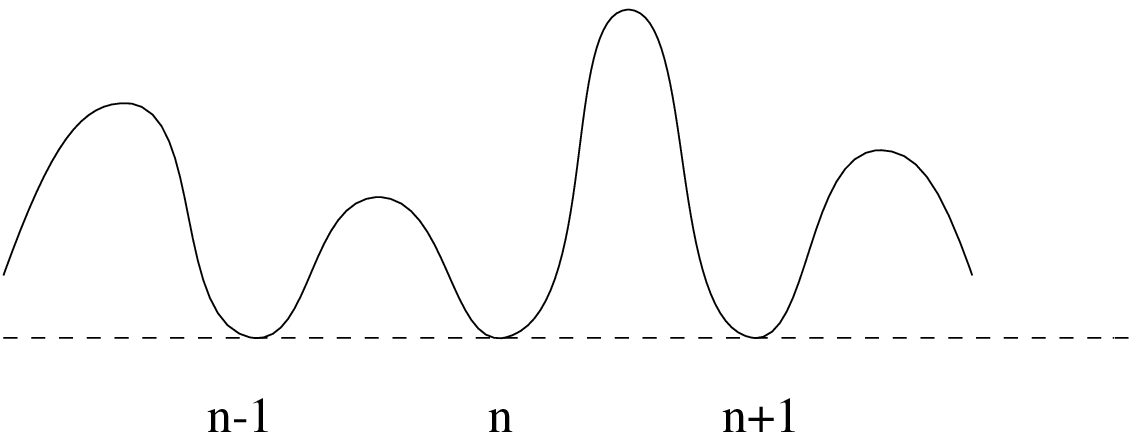, width=8cm}\\
(c)\\
          \caption{(a) Unsymmetric random potential, (b) random
            potential with equally high mountains, and (c) random
            potential with equally deep valleys.}
      \label{randompot}
    \end{center}
  \end{figure}

The configurations of a driven diffusive system with homogenous driving
force are fully characterized by a single conserved scalar field,
i.e., the local particle density $ n({\bf r},t) $. As the
particle number is conserved the order parameter for both phase
transitions is the deviation of the actual density $ n({\bf r},t) $
from its uniform average $ n_{0} $:
\begin{equation}
   \label{ord}
    s({\bf r},t) = n({\bf r},t) - n_{0} \; .
 \end{equation}
This fluctuating variable satisfies a continuity equation
\begin{equation}
  \label{konti}
  \dot{s} \;\; + \;\; \nabla \cdot {\bf j} \;\; = \;\; 0 
\end{equation}
where the current density $ {\bf j} $ consists of a deterministic and
a random part. The resulting stochastic differential equation is
a Langevin equation. First we model the deterministic
part. Due to the anisotropy of the system caused by the driving force,
$ {\bf j} $ shows different symmetries longitudinal and transverse to
its direction. Henceforth the indices $ \parallel
$ and $ \perp $ denote the spatial direction longitudinal to the
driving force and the $ (d-1) $--dimensional subspace transverse to
it, respectively. The transverse current is according to 
\begin{equation}
  \label{chempotj}
  {\bf j}_{\perp} = - \lambda \nabla_{\perp} \mu_{\perp}
\end{equation}
caused by a chemical potential $ \mu_{\perp} $ with $ \lambda $ being
a kinetic coefficient. As no direction is selected in the $ (d-1)
$--dimensional subspace the transverse current $ {\bf j}_{\perp} $ is
isotropic and therefore a vector of the form
\begin{equation}
  \label{jperp}
  {\bf j}_{\perp} = \nabla_{\perp} f(s,\Delta_{\perp},\nabla_{\parallel})
\end{equation}
where the Laplacean $ \Delta_{\perp} $ as argument denotes an even number of
$ \nabla_{\perp} $--operators in every term of $ f $. An expansion of
$ f $ with respect to $ s $ and gradient--operators yields, up to
higher order terms 
\begin{equation}
  \label{jperp2}
  {\bf j}_{\perp} = \nabla_{\perp}(s + s^{2} + 
  \Delta_{\perp} s + \nabla_{\parallel} s + 
  \Delta_{\parallel} s ) \; ,
\end{equation}
where here and in the following equations coefficients are suppressed for
the sake of simplicity.\\
The longitudinal current, however, possesses no symmetry, due to the
driving force. As a scalar it can be written as
\begin{equation}
  \label{jparallel}
  j_{\parallel} = g(s,\nabla_{\parallel},\Delta_{\perp})
\end{equation}
where the isotropy of the transverse subspace is again taken into
account by the even number of $ \nabla_{\perp} $--operators. Its
expansion up to higher order terms reads
\begin{equation}
  \label{jparallel2}
  j_{\parallel} = c + s +
   s^{2} + \nabla_{\parallel} s +
  \nabla_{\parallel}^{2} s + \nabla_{\parallel}^{3} s +
  \nabla_{\parallel} s^{2} + \Delta_{\perp} s +
  \nabla_{\parallel} \Delta_{\perp} s \; .  
\end{equation}
While some terms in this equation originate in a chemical potential $ \mu_{\parallel} $, according to 
\begin{equation}
  \label{chempot}
  j_{\parallel} = - \lambda \nabla_{\parallel} \mu_{\parallel} \; ,
\end{equation}
others are due to the driving force $ {\bf E} $, which at least produces
the terms proportional to $c$, $s$, $s^{2}$, and $\Delta_{\perp} s$
and which, in principle, also contributes to all terms of the
longitudinal current 
in (\ref{jparallel2}). Here, the constant c is the homogenous
part of the current, and $s^{2}$ is the first nonlinear term of the
longitudinal current and so the leading
nonlinearity of the problem.\\
Note that a repeated coarse graining of a microscopic model typically
generates anisotropic transport coefficients. Building here a model
directly on a mesoscopic length scale we must take that into
consideration. Thus, although containing the same terms, $
\mu_{\parallel} $ and $ \mu_{\perp} $ in general have different
coefficients, due to the anisotropy. From a technical point of view
anisotropic transport coefficients are required to make the model
renormalizable.\\
Following microscopic models and simulations
\cite{KLS1,KLS2,Marro,Leung1,Leung2} we restrict ourselves to such
models which additionally hold a CP (charge and parity)--symmetry,
i.e. in which the Langevin equation is invariant under the
transformation
\begin{equation}
  \label{cp}
  s(r_{\parallel},{\bf r}_{\perp},t) \rightarrow - s(-r_{\parallel},{\bf
  r}_{\perp},t) \; .
\end{equation}
In microscopic lattice models which also form the basis for Monte
Carlo simulations of driven diffusive systems the PC--symmetry
corresponds to the particle--hole symmetry in the half--filled
system.\\
In models with PC--symmetry all terms from (\ref{jperp2}) and
(\ref{jparallel2}) vanish whose sum of $ s $-- and $
\nabla_{\parallel} $--factors in the Langevin equation is even.\\
The stochastic part of the Langevin equation $ \zeta = - \nabla \cdot
{\bf j}_{L} $ reflects a random current $ {\bf j}_{L} $ that
summarizes the fast 
microscopic degrees of freedom (local in time in our Markovian
continuum description) and the effects of the quenched
disorder of the medium. After writing down the general form of the
Langevin equation 
we show how to model the various possibilities of quenched disorder
and which parts of the noise are relevant in the renormalization group
sense. Note that the noise considered here is conserving, i.e., it satisfies
the continuity equation. We mention that we have also analyzed driven
diffusive systems with nonconserving noise that is caused by random
particle sources \cite{BJ2,BJ4}.\\
Taking anisotropic transport coefficients into account we obtain the
Langevin equation
\begin{equation}
  \label{langevin}
  \dot{s} =\lambda [\Delta_{\perp}(\tau_{\perp} - \kappa_{\perp}
  \Delta_{\perp}) + \rho \Delta_{\parallel} ( \tau_{\parallel} -
  \kappa_{\parallel}\Delta_{\parallel}) -
  \kappa\Delta_{\perp}\Delta_{\parallel}] s + \frac{1}{2}\lambda g
  \nabla_{\parallel} s^{2} - \nabla \cdot {\bf j}_{L} \; . 
\end{equation}
This constitutes the fundamental equation for driven diffusive systems
with homogenous driving force, attractive interactions, CP--symmetry
and conserving noise. It contains all terms which are relevant in the
renormalization group sense for the two phase transitions and the
noncritical disordered phase, but in each case it still contains
irrelevant terms which must be eliminated by a dimensional analysis.\\
We proceed by investigating the influence of the quenched disorder in
detail. In a microscopic driven lattice gas model the quenched
disorder is modelled by random potential barriers between the
sites. There are three possible realizations of their randomness which
is depicted for one dimensional systems in Fig. \ref{randompot}. 
\begin{enumerate}
\item[I] The particles are in randomly deep potential valleys that are
  separated by randomly high potential mountains (Fig. \ref{randompot}
  (a)).
\item[II] The potential valleys are randomly deep, the potential
  mountains are equally high (Fig. \ref{randompot} (b)).
\item[III] The potential mountains are randomly high, but the
  potential valleys are equally deep (Fig. \ref{randompot} (c)).
\end{enumerate}
The homogeneous driving force tilts this landscape in the longitudinal
direction by an angle that depends on its strength. Thus in this
direction the symmetry of the random potential in the realizations II
and III is broken, whereas the symmetries in the transverse subspace
are not affected. Hence in the driving force direction the random potential is
unsymmetric in all realizations.\\
In the microscopic lattice gas model particles can only jump to
neighbouring unoccupied sites with jump rates that depend on the
external driving force, on the energetic situation of the particles due to
their attractive interaction, and on the locally random height of the
potential barriers between the sites. Performing a continuum limit of
the microscopic model one can show that, first, the main effect of the
quenched disorder results in a time--independent random current with
zero mean and 
Gaussian fluctuations and that, second, the three different realizations
of the random potential produce three different types of such a random
current $ \zeta_{d}({\bf r}) $.\\
In the unsymmetric case (realization I) the correlations of the random
current are given by
\begin{eqnarray}
  \label{stromerhalt}
  < {\bf \zeta}_{d}({\bf r})> & = & 0 \\
  < {\bf \zeta}_{d}({\bf r}) \otimes {\bf \zeta}_{d}({{\bf r'}}) > & = & 2\lambda^{2}
  \delta ({\bf r} - {\bf r'}) [\sigma {\bf e}_{\parallel}
  \otimes {\bf e}_{\parallel} + \gamma ({\bf 1} - {\bf e}_{\parallel}
  \otimes {\bf e}_{\parallel})] \; .\nonumber
\end{eqnarray}
Thus the correlations of the noise force $ \nabla \cdot \zeta_{d} $
read
\begin{eqnarray}
  \label{erhalt}
  < \nabla \cdot \zeta_{d}({\bf r})> & = & 0 \\
  < \nabla \cdot \zeta_{d}({\bf r}) \nabla \cdot \zeta({{\bf r'}}) > &
  = & -2\lambda^{2} 
  (\gamma \Delta_{\perp} + \sigma \Delta_{\parallel}) \delta ({\bf r} - {\bf
  r'}) \;\; .\nonumber
\end{eqnarray}
By a suitable scale change of $ s $ the kinetic coefficient $ \lambda
$ in equation(\ref{erhalt}) is the same as in
equation(\ref{langevin}). The anisotropy of the system due to the
driving force is taken into account here by the transport coefficients
$ \gamma $ and $ \sigma $.\\
In realization II a particle at a given site sees equally high
potential barriers in all transverse directions. As a consequence of
this symmetry the random current in the transverse subspace vanishes
in the lowest order and starts with a local gradient. The correlations
of the random current are here given by
\begin{equation}
  \label{brauschen}
  < \nabla \cdot \zeta({\bf r}) \nabla \cdot \zeta({{\bf r'}}) > \; = \; -2\lambda^{2}
  (-\alpha \Delta_{\perp}^{2} + \sigma \Delta_{\parallel}) \delta ({\bf r} - {\bf
  r'}) \; .
\end{equation}
As the symmetry of the random potential is broken by the driving force
in the longitudinal direction the longitudinal random current behaves
as in the unsymmetric case.\\
In realization III a potential barrier looks the same from both sides
in the transverse subspace. This is why the transverse random current
vanishes totally. Being only longitudinal the random current has the
correlations 
\begin{equation}
  \label{trauschen}
  < \nabla \cdot \zeta({\bf r}) \nabla \cdot \zeta({{\bf r'}}) > \; = \; -2\lambda^{2}
  \sigma \Delta_{\parallel} \delta ({\bf r} - {\bf
  r'}) \; .
\end{equation}
Since the continuum limit of the microscopic model is not rigorous, it
might be possible that even in the symmetric realizations II and III transverse
noise terms with $ \alpha \neq 0 $ and $ \gamma \neq 0 $,
respectively, are generated by the coarse graining procedure. This is
due to the fact that the symmetry of the random potential is
microscopic and concerns only a single site or bond. As we will see
later each type of noise nevertheless leads to different critical
behaviour governed by different stable fixed points with finite
regions of attraction. We defer the discussion 
of the possible additional transverse noise terms until the end of the
Sections 3.2 and 3.3, respectively.\\
By dimensional considerations (power counting) one can prove that each
of the equations (\ref{erhalt}) -- (\ref{trauschen}) contains all
relevant noise terms in the renormalization group sense for the
respective type of the random potential. Also deviations from
the local Gaussian nature of the random current $ \zeta $ are
irrelevant for the critical and long--time properties. Even the
microscopic part $ \zeta_{m} $ of the random current summarizing the
fast microscopic degrees of freedom is irrelevant if $ \zeta_{d} \neq
0 $. It should be remarked that $ \zeta_{m} $ plays the role of a
dangerous irrelevant field for the calculation of correlation
functions with frequencies $ \omega \neq 0 $. Without $ \zeta_{m} $
correlation functions show only a "central peak" at $ \omega = 0 $.

\section{Transverse Phase Transition}
In the following all models are analyzed with the help of renormalized
field theory. This method has successfully been applied to all driven
diffusive models investigated so far.\\
The three different realizations of the random potential described in
the last section actually lead to three different models for the
transverse phase transition and have to be treated separately.
\subsection{Unsymmetric Random Potential}
The model for a driven diffusive system with frozen random unsymmetric
potential is based on the equations (\ref{langevin}) and
(\ref{erhalt}).\\
To set up a renormalized field theory, it is convenient to recast the
model in terms of a dynamic functional \cite{J0,DeD,J1,J2,BJW,DeDP}
\begin{eqnarray}
  \label{udynfunktional}
  {\cal J}[s,\tilde{s}] & \hspace{-9 pt} =  &\hspace{-11 pt} \int\!\! d^{d}r \left\{\int\!\! dt [\tilde{s}\left (\dot{s} + \lambda
  (\Delta_{\perp}(\kappa_{\perp}\Delta_{\perp} - \tau_{\perp}) + \rho
  \Delta_{\parallel}(\kappa_{\parallel}\Delta_{\parallel} -
  \tau_{\parallel}) +\kappa \Delta_{\perp} \Delta_{\parallel}) s\right
  )\right.\nonumber\\
 &  & \hspace{30 pt}\left.  + \frac{1}{2} \lambda g
  (\nabla_{\parallel} \tilde{s}) s^{2} ]
   - \gamma \left[\lambda \int\!\! dt \nabla_{\perp} \tilde{s}\right]^{2}
   -\sigma \left[\lambda \int\!\! dt \nabla_{\parallel}
  \tilde{s}\right]^{2} \right\} \; ,
\end{eqnarray}
where $ \tilde{s}({\bf r},t) $ is a Martin--Siggia--Rose \cite{MSR}
response field. Correlation and response functions can now be
expressed as functional averages with weight $ \exp(-{\cal J}) $.\\
But for the description of the transverse phase transition this
dynamic functional still contains irrelevant terms in the
renormalization group sense. Now these terms are determined by a
dimensional analysis.\\
The transverse phase transition is characterized by finite $
\tau_{\parallel} $ and $ \tau_{\perp} \rightarrow 0 $. We introduce a
scale $ \mu^{2} $ for small $ \tau_{\perp} $. Then $ \mu^{-1} $ is a
convenient length scale. Since $ \tau_{\perp} $ tends to 0 at the
transverse phase transition, the leading term in the transverse direction is
proportional to $ \tilde{s} \lambda \Delta_{\perp} \kappa_{\perp}
\Delta_{\perp} s $, whereas in the longitudinal direction the leading
gradient term is proportional to $ \tilde{s} \lambda \rho
\Delta_{\parallel} \tau_{\parallel} s $. The comparison of the leading
gradient terms demonstrates that $ \Delta_{\parallel} $ scales as $
\Delta_{\perp}^{2} $. For longitudinal and transverse
length scales, this implies
\begin{equation}
  \label{skalierung}
  r_{\perp} \sim \mu^{-1} \qquad r_{\parallel} \sim \mu^{-2} \; .
\end{equation}
Since the dynamic functional is dimensionless the dimensions of fields
and coupling constants are
\begin{eqnarray}
  \label{udimension}
  \lambda t \sim \mu^{-4}\qquad &\qquad s \sim \mu^{\frac{d-5}{2}}\qquad & \qquad\tilde{s} \sim
  \mu^{\frac{d+7}{2}} \nonumber\\
  \kappa_{\parallel} \sim \mu^{-4}\qquad &\qquad \kappa \sim \mu^{-2}\qquad &\qquad \sigma
  \sim \mu^{-2} \nonumber\\
  \gamma \sim \mu^{0}\qquad & \qquad g \sim \mu^{\frac{9-d}{2}} \; ,\qquad & \qquad
\end{eqnarray}
i.e. the coupling constants $ \kappa_{\parallel}$, $ \kappa $, and $
\sigma $ are irrelevant in the renormalization group sense. The
irrelevancy of $ \sigma $ indicates that for the transverse phase
transition in a unsymmetric random potential the noise in the driving
force direction is irrelevant. From the
dimension of the nonlinear coupling constant $ g $ we recognize 
\begin{equation}
  \label{ugrenzdim}
   d_{c} = 9 
\end{equation}
as upper critical dimension of this model, above which $ g $ is
irrelevant und below which $ g $ is relevant. The finite $
\tau_{\parallel} $ and the transverse coupling constants are absorbed
into lengths and fields by a suitable scale change. Thus, the
appropriate dynamic functional to describe the transverse phase
transition in a driven diffusive system with an unsymmetric random
potential is given by
\begin{eqnarray}
  \label{uJ}
  {\cal J}[s,\tilde{s}] & = & \int\!\!
  d^{d}r \left\{\int\!\! dt \left[\tilde{s} \dot{s} + \lambda \tilde{s}
  (\Delta_{\perp}(\Delta_{\perp} - \tau_{\perp}) - \rho
  \Delta_{\parallel}) s + \frac{\textstyle 1}{\textstyle 2} \lambda g
  (\nabla_{\parallel} \tilde{s}) s^{2} \right]\right. \nonumber\\
  & & \hspace{190 pt}\left. - \left[\lambda \int\!\! dt \nabla_{\perp} \tilde{s}\right]^{2}
    \right\} \; .
\end{eqnarray}
The dynamic functional has the following symmetries. The
isotropy in the transverse subspace and the CP--symmetry that reads
here
\begin{equation}
  \label{dyncp}
  r_{\parallel} \rightarrow -r_{\parallel} \qquad s \rightarrow -s
  \qquad \tilde{s} \rightarrow - \tilde{s} 
\end{equation}
were directly integrated in the model building. After averaging over
the quenched disorder, $ \cal J $ exhibits translational symmetry in
space and time. The invariance of $ \cal J $
under the longitudinal scale transformation
\begin{eqnarray}
  \label{uskalen}
  r_{\parallel}   \rightarrow  \beta r_{\parallel} & \qquad &
  r_{\perp}  \rightarrow   r_{\perp}\\
  \tilde{s}  \rightarrow  \beta^{-\frac{1}{2}} \tilde{s} & \qquad &
  s   \rightarrow  \beta^{-\frac{1}{2}} s \nonumber\\
  \rho  \rightarrow  \beta^{2} \rho & \qquad & g   \rightarrow 
  \beta^{\frac{3}{2}} g \nonumber
\end{eqnarray}
is of great importance, because the parameter combination $
g^{2}\rho^{-\frac{3}{2}} $ is found as appropriate variable of the
model being invariant under this transformation. In contrast to the
corresponding model without quenched disorder \cite{JS2} the dynamic
functional is here not invariant under a Galilei transformation.\\
To study the critical properties of the transverse phase transition we apply
standard renormalization group methods \cite{Amit,BJW,DeDP}. We use
dimensional regularization in $ d = 9 - \epsilon $ followed by minimal
subtraction. The one--line--irreducible vertex functions with
$\tilde{n} \; \tilde{s} $--legs and $ n \; s $--legs at wavevectors $
\{{\bf q}\} $ and frequencies $ \{\omega\} $ will be denoted by 
$ \Gamma_{\tilde{n},n}(\{{\bf q}\},\{\omega\}) $. Taking into
consideration the causality properties of the theory we only find $
\Gamma_{1,1} $ and $ \Gamma_{1,2} $ primitively divergent. 
The nontrivial diagrams contributing to $ \Gamma_{\tilde{n},n}(\{{\bf
  q}\},\{\omega\}) $ carry at least a factor $
q_{\parallel}^{\tilde{n}} $, because the interaction vertex is
conserving. The primitive divergences are multiplicatively absorbed
in a redefinition of the parameters 
\begin{equation}
  \label{ureparameter}
  \rho \rightarrow \bare{\rho}  =  Z_{\rho} \rho \; , \;
  g \rightarrow \bare{g}  =  \mu^{\frac{\epsilon}{2}} Z_{u} u \; .
\end{equation}
Here, in contrast to the ordered problem \cite{JS2}, the coupling
constant $ g $ has to be renormalized, due to the loss of Galilean
invariance in the disordered problem.\\
The elements of our perturbation expansion are the Gaussian propagator
\begin{equation}
  \label{uprop}
  G_{{\bf q},\omega} \; := \; < \tilde{s}_{{\bf q},\omega} s_{-{\bf
  r},-\omega}>_{0} \; = \; \frac{1}{i\omega + \lambda[{\bf
  q}_{\perp}^{2}({\bf q}_{\perp}^{2} + \tau_{\perp}) + \rho
  q_{\parallel}^{2}]} \; ,
\end{equation}
the Gaussian correlator
\begin{equation}
  \label{ukorr}
  C_{{\bf q},\omega} \; := \; < s_{{\bf q},\omega} s_{-{\bf q},-\omega}>_{0}
   \; = \; \frac{2 \lambda^{2}{\bf q}_{\perp}^{2} \delta(\omega)}{\omega^{2}
  + \lambda^{2}[{\bf q}_{\perp}^{2}({\bf q}_{\perp}^{2} + \tau_{\perp}) + \rho
  q_{\parallel}^{2}]^{2}}  \; ,
\end{equation}
and the conserving vertex $ v({\bf q}) = -i\lambda g q_{\parallel}
$. Their graphical representation is shown in the Appendix in
fig.(\ref{graphelem}).\\
As the correlator of the theory $ C_{{\bf q},\omega} $ is proportional
to $ \delta(\omega) $ (i.e. $ C_{{\bf q}}(t) $ is independent of time)
and any loop of a diagram contains at least one correlator because of
the structure of the vertex with two $ s $--legs and one $ \tilde{s}
$--leg and because of causality the integration over internal
frequencies cannot generate any divergences. Thus the divergent parts
of the vertex functions are only in the ($ \omega = 0 $)--parts. To
facilitate their calculation it is convenient to split the order
parameter
\begin{equation}
\label{aufspaltung}
  s({\bf r},t) = \varphi({\bf r}) + s^{\prime}({\bf r},t)
\end{equation}
into a time independent $ \varphi({\bf r}) $ and a time dependent $
s^{\prime}({\bf r},t) $ that contains no ($ \omega = 0
$)--parts. Since the noise is independent of time $ s^{\prime}({\bf
  r},t) $ relaxes deterministically. In the limit of long times we
obtain
\begin{equation}
\label{tlimes}
  s({\bf r},t \rightarrow \infty) = \varphi({\bf r}) \; .
\end{equation}
Together with 
\begin{equation}
  \label{quasitrans}
\tilde{\varphi}({\bf r}) = \lambda \int dt \tilde{s}({\bf r},t)
\end{equation}
the dynamic functional $ \cal{J} $ reduces to a quasi--static (frozen)
Hamiltonian 
\begin{equation}
   \label{quasihamil}
   {\cal H}[\varphi,\tilde{\varphi}] =  \int\!\!
  d^{d}r \left\{\tilde{\varphi}
  (\Delta_{\perp}(\Delta_{\perp} - \tau_{\perp}) - \rho
  \Delta_{\parallel}) \varphi + \frac{1}{2} g
  (\nabla_{\parallel} \tilde{\varphi}) \varphi^{2} + \tilde{\varphi}
  \Delta_{\perp} \tilde{\varphi}  \right\} \; .
 \end{equation}
The frozen Hamiltonian generates all the zero--frequency parts of the
vertex functions if causality is included in the graphical rules of
perturbation theory.\\
In the quasi--static model the Gaussian propagator and correlator are
only dependent on wavevectors and read
 \begin{eqnarray}
  \label{quasipropkorr}
   G_{{\bf q}} & = & \frac{1}{{\bf
  q}_{\perp}^{2}({\bf q}_{\perp}^{2} + \tau_{\perp}) + \rho
  q_{\parallel}^{2}} \\ 
  C_{{\bf q}}  & = &\frac{2 {\bf q}_{\perp}^{2} }{[{\bf
  q}_{\perp}^{2}({\bf q}_{\perp}^{2}
 + \tau_{\perp}) + \rho  q_{\parallel}^{2}]^{2}}  \; .\nonumber
\end{eqnarray}
In the quasi--static model we calculate the primitively divergent
vertex functions in two--loop order. While the one--loop calculation
is easy to perform analytically, technical difficulties have to be
overcome for the two--loop diagrams with mixed $ {\bf q}^{4}
$--propagators and --correlators. These two--loop diagrams have here
been solved by a new technique where a special inverse Mellin
transformation is used to factorize the denominators of some
propagators and correlators, respectively. Thus, the $\epsilon^{2}
$--poles and the simple $ \epsilon $--poles can be extracted. Their
coefficients are then given by parameter integrations over paths in the
complex plane. Whereas the coefficients of the $\epsilon^{2}
$--poles can be analytically calculated,  those of the simple $
\epsilon $--poles 
have to be computed numerically. The details of the two--loop
calculation are shown in the Appendix.\\
In a two--loop calculation using dimensional regularization we obtain
the vertex functions in an expansion in $ \epsilon = d_{c} - d $ and $
{\bf q} $ 
 \begin{eqnarray}
  \label{uvertex}
\bare{\Gamma}_{1,1}({\bf q})\!\! &\!\! =\!\! & \!\!{\bf
  q}_{\perp}^{2}({\bf q}_{\perp}^{2} + \tau_{\perp}) + \bare{\rho}
  q_{\parallel}^{2} + \frac{2}{3} \frac{A_{\epsilon}}{\epsilon}
  \frac{\bare{g}^{2}}{\bare{\rho}^{\frac{1}{2}}} q_{\parallel}^{2}
  \tau_{\perp}^{-\frac{\epsilon}{2}} - \frac{2}{9}
 \frac{A_{\epsilon}^{2}}{\epsilon^{2}}
  \frac{\bare{g}^{4}}{\bare{\rho}^{2}} q_{\parallel}^{2}
  \tau_{\perp}^{-\epsilon}(1 - 0.2158 \epsilon) \nonumber\\
\bare{\Gamma}_{1,2}({\bf q})\!\! &\!\! =\!\! &\!\! i \bare{g} q_{\parallel} - i \frac{1}{6}
  \frac{A_{\epsilon}}{\epsilon}
  \frac{\bare{g}^{3}}{\bare{\rho}^{\frac{3}{2}}} q_{\parallel}
  \tau_{\perp}^{-\frac{\epsilon}{2}} + \frac{1}{8} i 
 \frac{A_{\epsilon}^{2}}{\epsilon^{2}}
  \frac{\bare{g}^{5}}{\bare{\rho}^{3}} q_{\parallel}
  \tau_{\perp}^{-\epsilon}(1 - 0.2154 \epsilon) \; , 
 \end{eqnarray}
where bare unrenormalized parameters are indicated by a superscript ``$\;
\bare{} $ " above the symbol. $ A_{\epsilon} = \frac{1}{(2\pi)^{d}}
{\cal O}_{d-1} \Gamma(1 + 
  \frac{\epsilon}{2}) \Gamma(\frac{5 - \epsilon}{2})
  \Gamma(\frac{1}{2}) $ is a suitably chosen constant with $ {\cal
    O}_{d-1} $ being the surface of the $ (d-1) $--dimensional unit
  sphere and $ \Gamma(z) $ denoting Euler's $ \Gamma $--function. The
  $ Z $--factors defined in equation(\ref{ureparameter}) are in minimal
  subtraction
\begin{eqnarray}
  \label{uZ}
  Z_{\rho} & = & 1 - \frac{2}{3 \epsilon}  A_{\epsilon}
  \frac{u^{2}}{\rho^{\frac{3}{2}}} - \frac{2}{9
  \epsilon^{2}}A_{\epsilon}^{2}\frac{u^{4}}{\rho^{3}}(1 + 0.2158 \epsilon) + {\cal O}(u^{6}) \\
  Z_{u}    & = & 1 + \frac{1}{6 \epsilon}  A_{\epsilon}
  \frac{u^{2}}{\rho^{\frac{3}{2}}} + \frac{1}{8
  \epsilon^{2}}A_{\epsilon}^{2}\frac{u^{4}}{\rho^{3}}(1 + 0.2154 \epsilon) + {\cal O}(u^{6})  \; .\nonumber
\end{eqnarray}
We recognize that the perturbation expansion is organized in powers of
the dimensionless renormalized parameter combination $ v :=
A_{\epsilon} u^{2}\rho^{-\frac{3}{2}} $ that was already found as
invariant variable under the longitudinal scale transformation
(equation(\ref{uskalen})).\\
With the renormalizations at hand we are in a position to determine
the critical behaviour of the vertex functions. We use the fact that the
unrenormalized theory is independent of the momentum scale $ \mu
$. This leads to the renormalization group equation
\begin{equation}
  \label{rgg}
  \left [\beta_{v} \frac{\partial}{\partial v} + \rho \zeta \frac{\partial}{\partial
  \rho} + \mu \frac{\partial}{\partial \mu}\right ] \Gamma_{\tilde{n},n}
  (\{q_{\parallel},{\bf q}_{\perp}\},\tau_{\perp},v,\rho,\mu) = 0 \; .
\end{equation}
The Wilson parameter functions, being only dependent on $ v $, are given
by
\begin{eqnarray}
  \label{uwilson}
  \beta_{v} & := & \left. \mu \frac{\partial v}{\partial \mu}\right
  |_{o} = -v (\epsilon - \frac{4}{3} v - 0.252 v^{2} + {\cal
  O}(v^{3}))\\  
  \zeta & := & \left. \mu \frac{\partial \ln \rho}{\partial
  \mu}\right |_{o} = - \frac{2}{3} v - 0.0959 v^{2} + {\cal O}(v^{3})
  \; ,\nonumber 
\end{eqnarray}
where the derivatives are calculated at fixed bare parameters. As a
linear partial differential equation the renormalization group
equation is solvable by the method of characteristics with the result
\begin{equation}
  \label{urggresult}
  \Gamma_{\tilde{n},n}
  (\{q_{\parallel},{\bf q}_{\perp}\},\tau_{\perp},v,\rho,\mu) = \Gamma_{\tilde{n},n}
  (\{q_{\parallel},{\bf
  q}_{\perp}\},\tau_{\perp},\bar{v}(l),\bar{\rho}(l),\bar{\mu}(l)) \; .
\end{equation}
The trajectories $ \bar{v}(l) $, $ \bar{\rho}(l) $, and $ \bar{\mu}(l)
$ are solutions of the flow equations
\begin{equation}
  \label{ufluss}
  l \frac{d}{dl} \bar{v}(l) = \beta_{v}(\bar{v}(l)) \qquad l
  \frac{d}{dl} \ln \bar{\rho}(l) = \zeta(\bar{v}(l)) \qquad
  l \frac{d}{dl} \bar{\mu}(l) = \bar{\mu}(l)
\end{equation}
with the flow parameter $ l $ and the initial conditions
\begin{equation}
  \label{uanfang}
  \bar{v}(l=1) = v \qquad \bar{\rho}(l=1) = \rho \qquad \bar{\mu}(l=1)
  = \mu \; .
\end{equation}
The third flow equation obviously has the solution $ \bar{\mu}(l) =
\mu l $.\\
With the help of characteristics the critical asymptotic region of
small $ \tau_{\perp} $ and $ q $ can be mapped onto uncritical
regions. In the scaling limit $ l \ll 1 $ corresponding to 
\begin{equation}
  \label{skalenlimes}
  \left |\frac{q_{\parallel}}{\mu^{2}} \right | \ll 1 \qquad
  \left |\frac{{\bf q}_{\perp}}{\mu} \right | \ll 1 \qquad
  \left |\frac{\tau_{\perp}}{\mu^{2}} \right | \ll 1
\end{equation}
the flow of $ \bar{v}(l) $ is controlled by stable zeroes of $
\beta_{v} $. For $ \epsilon > 0 $, there is a nontrivial infrared
stable fixed point
\begin{equation}
  \label{ufixpunkt}
  v_{*} \; = \; \frac{\textstyle 3}{\textstyle 4} (\epsilon - 0.141 \epsilon^{2} + {\cal
   O}(\epsilon^{3}))   \; .
\end{equation}
The other fixed point $ v_{*} = 0 $ being Gaussian is stable only if $
\epsilon < 0 $.\\
From the nontrivial fixed point value $ \zeta(v_{*}) $ we define the anomaly
exponent $ \eta := -\frac{1}{2} \zeta(v_{*}) $ that is to two--loop
order
\begin{equation}
  \label{ueta}
  \eta = \frac{\textstyle 1}{\textstyle 4} \epsilon (1 -
 0.302 \; \frac{\textstyle \epsilon}{\textstyle 9} + {\cal
  O}(\epsilon^{2})) \; .
\end{equation}
At the fixed point the solution of the second flow equation of
equation (\ref{ufluss}) is 
\begin{equation}
  \label{rhofluss}
  \bar{\rho}(l) = \rho l^{-2 \eta} \; .
\end{equation}
The results found for the vertex functions in the quasi--static model
can be directly transferred into the dynamic model because the time scale
$ \lambda $ needs no renormalization.\\
The longitudinal length scale transformation according to equation
(\ref{uskalen}), dimensional analysis and the renormalization group equation
are now combined to derive the asymptotic critical scaling form of the
vertex functions in the dynamic model at the transverse phase
transition
\begin{eqnarray}
  \label{uskalkrit}
  \Gamma_{\tilde{n},n}
  (\{q_{\parallel},{\bf q}_{\perp},\omega\},\tau_{\perp},v_{*},\lambda,\rho,\mu)
  & = & \nonumber\\
  & & \hspace{-170 pt}l^{-\frac{1}{2} \eta (\tilde{n} + n -2)
  -\frac{1}{2} \tilde{n} (d+7) - \frac{1}{2} n(d-5) + d + 5}  \Gamma_{\tilde{n},n}
  \left(\left\{\frac{\textstyle q_{\parallel}}{\textstyle l^{2+\eta}},\frac{\textstyle {\bf q}_{\perp}}{\textstyle l},\frac{\textstyle \omega}{\textstyle l^{4}}\right\},
\frac{\textstyle \tau_{\perp}}{\textstyle l^{2}},v_{*},\lambda,\rho,\mu\right)
\; .
\end{eqnarray}
This equation implies that, below the upper critical dimension $ d_{c} =
9 $, anomalous scaling behaviour only occurs in the direction of the
driving force and is completely characterized by the anomaly exponent
$ \eta $ from (\ref{ueta}) that is positive. These results are in
analogy to what was found in the driven diffusive systems investigated
so far \cite{JS1,JS2,BJ1,BJ2}.\\
To illustrate the importance of the positive anomaly exponent $ \eta $
we especially investigate the density response function $ \chi({\bf
  q},t) $ which is the Fourier transform of $ \Gamma_{1,1}^{-1}({\bf
  q},\omega) $. Choosing the flow parameter $ l =
\omega^{\frac{1}{4}} \ll 1 $, we derive the scaling form of the density
response function $ \chi({\bf q},t):= <s({\bf q},t) \tilde{s}(-{\bf
  q},0)> $ from equation (\ref{uskalkrit}):
\begin{equation}
  \label{usus}
  \chi({\bf q},t) \; = \; f\left(q_{\parallel}^{2} t^{1+\frac{1}{2}
  \eta},q_{\perp}^{2} t^{\frac{1}{2}}\right) \; .
\end{equation}
From that we conclude that for long times typical longitudinal
length--squares scale with time as 
\begin{equation}
  \label{ulangskal}
  < r_{\parallel}^{2} > \; \sim \; t^{1+\frac{1}{2} \eta} \; .
\end{equation}
The positivity of $ \eta = \frac{\textstyle 1}{\textstyle 4} \epsilon (1 -
 0.302 \; \frac{\textstyle \epsilon}{\textstyle 9} + {\cal
  O}(\epsilon^{2})) $ means that in systems with spatial dimensions $ d
  < d_{c} = 9 $ fluctuations spread faster than diffusively in the
  driving force direction. This superdiffusion was also observed in
  all driven diffusive systems analyzed up to now
  \cite{JS1,JS2,BJ1,BJ2}.\\
A comparison with the corresponding model for the transverse phase
  transition without quenched disorder, having an upper critical
  dimension $ d_{c} = 5 $, shows that the phenomenon of superdiffusion
  occurs here over a much greater dimensional range. As the system
  without quenched disorder behaves normally diffusive for $ d > 5 $
  it follows that quenched disorder is the reason for superdiffusion
  in the dimensional interval $ 5 < d < 9 $.\\
The analogous comparison of the models with and without disorder for
  the noncritical region \cite{JS1,BJ1} came to an analogous result.\\
Another consequence from the scaling form of the density response
  function is that typical transverse length--squares scale as $ <
  r_{\perp}^{2} > \; \sim \; t^{\frac{1}{2}} $ for long times,
  i.e. subdiffusively. This behaviour corresponds to the naive
  dynamical exponent $ z = 4 $ in model B (in the nomenclature of
  Halperin and Hohenberg \cite{HH}) and has been expected, for the
  system here is critical with respect to the transverse directions
  and no renormalizations are necessary in this subspace.\\
The extrapolation of the anomaly--exponent $ \eta $ into low
 dimensions is difficult in view of the high upper critical
 dimension $ d_{c} = 9 $. Naively, we can simply set $ \epsilon = d_{c} -d = 
  6 $ in Equation (\ref{ueta}) to predict $ \eta $ f.e. in a
 3--dimensional system, resulting in an estimate
\begin{equation}
  \label{ueta32}
  \eta(d=3) \; = \; 1.20 \; .
\end{equation}
Although $ \epsilon $ has been considered as small quantity the two
loop--correction with respect to the one--loop result is merely 20\%,
even at $ \epsilon = 6 $. Due to these small corrections we expect
that equation (\ref{ueta}), even though it is strictly valid only near
$ d_{c} = 9 $, 
produces good approximations for $ \eta $ also in low dimensions.
\subsection{Random Potential With Equally High Potential Mountains}
This model is analyzed in analogy to the model with unsymmetric random
potential of the last section. The quenched disorder according to
Equation (\ref{brauschen}) resulting from
the random potential with equally high potential mountains leads, in
analogy to Equation (\ref{udynfunktional}), to
the dynamic functional
\begin{eqnarray}
  \label{bdynfunkt}
  {\cal J}[s,\tilde{s}] & \hspace{-9 pt} =  &\hspace{-11 pt} \int\!\! d^{d}r \left\{\int\!\! dt [\tilde{s}\left (\dot{s} + \lambda
  (\Delta_{\perp}(\kappa_{\perp}\Delta_{\perp} - \tau_{\perp}) + \rho
  \Delta_{\parallel}(\kappa_{\parallel}\Delta_{\parallel} -
  \tau_{\parallel}) +\kappa \Delta_{\perp} \Delta_{\parallel}) s\right
  )\right.\nonumber\\
 &  & \hspace{30 pt}\left.  + \frac{1}{2} \lambda g
  (\nabla_{\parallel} \tilde{s}) s^{2} ]
   - \alpha \left[\lambda \int\!\! dt \Delta_{\perp} \tilde{s}\right]^{2}
   -\sigma \left[\lambda \int\!\! dt \nabla_{\parallel}
  \tilde{s}\right]^{2} \right\} \; ,
\end{eqnarray}
that still contains irrelevant terms to be eliminated.\\
Transverse and longitudinal lengths scale as before $ r_{\perp} \sim \mu^{-1},\;
r_{\parallel} \sim \mu^{-2} $ . Since the dynamic functional is
dimensionless the dimension of fields and coupling constants are 
\begin{eqnarray}
  \label{bdimension}
  \lambda t \sim \mu^{-4}\qquad &\qquad s \sim \mu^{\frac{d-3}{2}}\qquad & \qquad\tilde{s} \sim
  \mu^{\frac{d+5}{2}} \nonumber\\
  \kappa_{\parallel} \sim \mu^{-4}\qquad &\qquad \kappa \sim \mu^{-2}\qquad & \qquad\sigma
  \sim \mu^{0} \nonumber\\
  \alpha \sim \mu^{0}\qquad &\qquad  g \sim \mu^{\frac{7-d}{2}}\qquad & \qquad\qquad\; .
\end{eqnarray}
Thus, the coupling constants $ \kappa_{\parallel} $ and $ \kappa $ are
again irrelevant in the renormalization group sense, but $ \sigma $ is
not. This signifies that in the given random potential both longitudinal and
transverse noise are relevant. The dimension of the nonlinear coupling
$ g $ shows that
\begin{equation}
  \label{bgrenzdim}
  d_{c} = 7 
\end{equation}
is the upper critical dimension of this model. The finite
$ \tau_{\parallel} $ and the transverse coupling constants $ \alpha $
 and $ \kappa_{\perp} $ are again absorbed by a suitable scale
 change. Thus, the appropriate dynamic functional for the transverse
 phase transition in the given random potential is
\begin{eqnarray}
  \label{bJ}
  {\cal J}[s,\tilde{s}] & = & \int\!\!
  d^{d}r \left\{\int\!\! dt \left[\tilde{s} \dot{s} + \lambda \tilde{s}
  (\Delta_{\perp}(\Delta_{\perp} - \tau_{\perp}) - \rho
  \Delta_{\parallel}) s + \frac{\textstyle 1}{\textstyle 2} \lambda g
  (\nabla_{\parallel} \tilde{s}) s^{2} \right]\right. \nonumber\\
  & & \hspace{100 pt}\left. - \left[\lambda \int\!\! dt \Delta_{\perp} \tilde{s}\right]^{2}
                            - \sigma \left[\lambda \int\!\! dt
  \nabla_{\parallel} \tilde{s} \right]^{2} \right\} 
 \; .
\end{eqnarray}
This dynamic functional exhibits the same symmetries as the one in the
case of the unsymmetric random potential, except the longitudinal
scale transformation, where here the parameter $ \sigma $ is additionally
transformed according to $ \sigma \rightarrow \beta^{2} \sigma
$. In addition to the parameter combination $ g^{2} \rho^{-\frac{3}{2}}
$, $ \sigma \rho^{-1} $ is also an invariant variable of this model and
the perturbation expansion will be organized in powers of both
variables.\\ 
Due to the quenched disorder, the Gaussian correlator is again time
independent so that, for simplicity, we also here transform $ \cal{J}
$ into a quasi--static Hamiltonian via (\ref{tlimes}) and (\ref{quasitrans})
\begin{eqnarray}
   \label{bquasihamil}
   {\cal H}[\varphi,\tilde{\varphi}] & = & \int\!\!
  d^{d}r \left\{\tilde{\varphi}
  (\Delta_{\perp}(\Delta_{\perp} - \tau_{\perp}) - \rho
  \Delta_{\parallel}) \varphi + \frac{1}{2}  g
  (\nabla_{\parallel} \tilde{\varphi}) \varphi^{2} \right.\\
  & & \hspace{170 pt} \left. + \tilde{\varphi}
  (-\Delta_{\perp}^{2} + \sigma \Delta_{\parallel}) \tilde{\varphi}  \right\} \; .\nonumber
 \end{eqnarray}
In comparison with the former model, the vertex and the Gaussian
propagator remain the same (\ref{quasipropkorr}), whereas the Gaussian
correlator of the quasistatic model reads
\begin{equation}
  \label{bpropkorr}
 C_{{\bf q}}   = \frac{2({\bf q}_{\perp}^{4} + \sigma q_{\parallel}^{2}) }{[{\bf
  q}_{\perp}^{2}({\bf q}_{\perp}^{2}
 + \tau_{\perp}) + \rho  q_{\parallel}^{2}]^{2}}  \; .  
\end{equation}
By dimensional analysis taking into account the causality properties
we find the vertex functions $ \Gamma_{1,1} $, $ 
\Gamma_{2,0} $, and $ \Gamma_{1,2} $ primitively divergent.\\
Neglecting higher orders in $ q $ and $
\epsilon = d_{c} - d $, a one--loop calculation for these
vertex functions, with dimensional regularization, results in 
\begin{eqnarray}
  \label{bvertex}
  \bare{\Gamma}_{1,1}({\bf q}) & = & {\bf
  q}_{\perp}^{2}({\bf q}_{\perp}^{2} + \tau_{\perp}) + \bare{\rho}
  q_{\parallel}^{2} + \frac{B_{\epsilon}}{\epsilon}
  \frac{\bare{g}^{2}}{\bare{\rho}^{\frac{1}{2}}} q_{\parallel}^{2}
  \tau_{\perp}^{-\frac{\epsilon}{2}}\\
\bare{\Gamma}_{2,0}({\bf q}) & = & -2{\bf
  q}_{\perp}^{4} -2 \bare{\sigma}
  q_{\parallel}^{2} -\frac{1}{4} \frac{B_{\epsilon}}{\epsilon}
  \frac{\bare{g}^{2}}{\bare{\rho}^{\frac{1}{2}}} q_{\parallel}^{2}
  \tau_{\perp}^{-\frac{\epsilon}{2}}\left(5 + 2
  \frac{\bare{\sigma}}{\bare{\rho}} + \left(\frac{\bare{\sigma}}{\bare{\rho}}\right)^{2}\right)\nonumber\\
  \bare{\Gamma}_{1,2}({\bf q}) & = & i \bare{g} q_{\parallel} - i \frac{1}{4}
  \frac{B_{\epsilon}}{\epsilon}
  \frac{\bare{g}^{3}}{\bare{\rho}^{\frac{3}{2}}} q_{\parallel}
  \tau_{\perp}^{-\frac{\epsilon}{2}}\left(1 + \frac{\bare{\sigma}}{\bare{\rho}}\right) \; , \nonumber
\end{eqnarray}
where 
\begin{equation}
  \label{bgeofaktor}
  B_{\epsilon} = \frac{1}{(2\pi)^{d}} {\cal O}_{d-1} \Gamma(1 +
  \frac{\epsilon}{2}) \Gamma(\frac{3 - \epsilon}{2}) \Gamma(\frac{1}{2})
\end{equation}
is a suitably chosen $ \epsilon $--dependent factor. Notice that
although not performed a two--loop calculation could here also be done with the
method described in the Appendix.\\
The primitive divergences are absorbed in the redefinition of the
parameters
\begin{equation}
  \label{breparameter}
  \bare{\rho}  =  Z_{\rho} \rho, \;
  \bare{\sigma}  =  Z_{\sigma} \sigma, \;
  \bare{g}     = \mu^{\frac{\epsilon}{2}} Z_{u} u \; .
\end{equation}
In comparison to the latter model the coupling constant $
\bare{\sigma} $ has additionally to be renormalized. From Equation
(\ref{bvertex}) we easily obtain the $ Z $--Factors in minimal
subtraction
\begin{eqnarray}
  \label{bZ}
  Z_{\rho} & = & 1 - \frac{v}{\epsilon} + {\cal O}(v^{2}) \\
  Z_{\sigma} & = & 1 - \frac{1}{8} \frac{v}{\epsilon} \frac{1}{w} (5 +
  2w + w^{2}) + {\cal O}(v^{2}) \nonumber\\
  Z_{u} & = & 1 + \frac{1}{4} \frac{v}{\epsilon} (1 + w) + {\cal
  O}(v^{2}) \; ,\nonumber
\end{eqnarray}
expressed as functions of the dimensionless renormalized parameters $
v:= B_{\epsilon} u^{2} \rho^{-\frac{3}{2}} $ and $ w:=
\sigma \rho^{-1} $ that are invariant under the longitudinal scale
transformation.\\
In analogy to the former model we obtain the renormalization group
equation 
\begin{equation}
  \label{brgg}
  \left [\beta_{v} \frac{\partial}{\partial v} + \beta_{w}
  \frac{\partial}{\partial w}  +\rho \zeta \frac{\partial}{\partial
  \rho} + \mu \frac{\partial}{\partial \mu}\right ] \Gamma_{\tilde{n},n}
  (\{q_{\parallel},{\bf q}_{\perp}\},\tau_{\perp},v,w,\rho,\mu) = 0 \; 
\end{equation} 
with the parameter functions depending on $ v $ and $ w $
\begin{eqnarray}
  \label{bwilson}
  \beta_{v} & := & \left. \mu \frac{\partial v}{\partial \mu}\right |_{o} = -v
  \left[\epsilon - \frac{1}{2} v(4 + w) + {\cal O}(v^{2})\right]\\
\beta_{w} & := & \left. \mu \frac{\partial w}{\partial \mu}\right
  |_{o} = - \frac{1}{8} v [5 - 6w + w^{2} ] + {\cal O}(v^{2})
  \nonumber\\
  \zeta & := & \left. \mu \frac{\partial \ln \rho}{\partial
  \mu}\right |_{o} = - v + {\cal O}(v^{2}) \; . \nonumber
\end{eqnarray}
The associated characteristics are defined by 
\begin{eqnarray}
  \label{bfluss}
  l \frac{d}{dl} \bar{v}(l) = \beta_{v}(\bar{v}(l),\bar{w}(l)) &
  \qquad\qquad & l \frac{d}{dl} \ln \bar{\rho}(l) = \zeta(\bar{v}(l),\bar{w}(l)) \\
  l \frac{d}{dl} \bar{w}(l) = \beta_{w}(\bar{v}(l),\bar{w}(l)) &
  \qquad\qquad  & l \frac{d}{dl} \bar{\mu}(l) = \bar{\mu}(l) \nonumber
\end{eqnarray}
with the initial condition $ \bar{w}(l=1) = w $ and the initial
conditions from Equation (\ref{uanfang}). In the scaling limit $ l \ll
1 $, $ \bar{v}(l) $ and $ \bar{w}(l) $ flow to an infrared stable
fixed point $ (v_{*},w_{*}) $ given by the zeroes of $ \beta_{v}
$ and $ \beta_{w} $ with a positive gradient. The zeroes of $ \beta_{w}
$ in this order are
\begin{equation}
  \label{nullw}
  w_{1} \; = \; 1 \qquad \qquad w_{2} \; = \; 5 \; ,
\end{equation}
and those of $ \beta_{v} $
\begin{equation}
  \label{nullv}
  v_{1} \; = \; 0 \qquad \qquad v_{2} \; = \; \frac{2 \epsilon}{4 + w}
  \; .
\end{equation}
This yields the infrared stable fixed point
\begin{equation}
  \label{bfixpunkt}
  w_{*} = 1 + {\cal O}(\epsilon) \qquad 
        v_{*} = \frac{\textstyle 2}{\textstyle 5} \epsilon + {\cal O}(\epsilon^{2})   \; .
\end{equation} 
The domain of attraction of this fixed point can easily be recognized
as 
\begin{equation}
  \label{attrakt}
  v > 0 \qquad w < 5 + {\cal O}(\epsilon) \; .
\end{equation}
In the case $ w > 5 + {\cal O}(\epsilon) $ the critical behaviour of
the system is dominated by the degenerate fixed point 
\begin{equation}
  \label{entartfix}
  w_{*} = \infty \qquad v_{*} = 0 \; .
\end{equation}
Concerning this degenerate fixed point we remark the following:\\
i) For $ w \rightarrow \infty $ the transverse noise vanishes in
comparison to the longitudinal one. This can be seen easily by
substituting $ \rho q_{\parallel}^{2} \rightarrow q_{\parallel}^{2} $
and extracting $ \sigma \rho^{-1} = w $ from the correlator. Then the
remaining coefficient of the longitudinal part in the numerator of the
correlator is $ 1 $, the coefficient of the transverse part is $
w^{-1} $. At the degenerate fixed point, this system here behaves as the
model with a random potential with equally deep potential valleys and
whose noise is therefore given by Equation (\ref{trauschen}). This
model is analyzed in the next section, but we anticipate some results
concerning the degenerate fixed point. According to
Eq. (\ref{tfixpunkt}) the degenerate fixed point possesses a finite
fixed point value $ (v\cdot w)_{*} = \frac{8}{3} \epsilon +
{\cal O}(\epsilon^{2}) $. Although the deviations from normal
diffusive behaviour only appear in order two--loop, the positive $
\eta $ from Eq. (\ref{teta}) demonstrates the system to be also
superdiffusive at the degenerate fixed point.\\
ii) Both a ``normal" and a degenerate fixed point are also
observed in the noncritical model with quenched disorder
\cite{BJ1}. While here at the transverse phase transition the
situation of two fixed points appears in the random potential with
equally high mountains and the degenerate fixed point is described by
the model with the random potential with equally deep valleys, in the
noncritical region, however, the situation of a normal and a
degenerate fixed point occurs in the model with unsymmetric random
potential and the degenerate fixed point is described by both the model
with equally high mountains and equally deep valleys, for these latter
models are identical in the noncritical region.\\
Now we proceed to investigate the normal fixed point
(\ref{bfixpunkt}). For this fixed point the anomaly--exponent reads 
\begin{equation}
  \label{beta}
  \eta = \frac{\textstyle 1}{\textstyle 5} \epsilon + {\cal O}(\epsilon^{2}) 
  \; .
\end{equation}
At the fixed point, the second characteristic has the form 
\begin{equation}
  \label{brhofluss}
  \bar{\rho}(l) = \rho l^{-2 \eta} 
\end{equation}
in analogy to Equation (\ref{rhofluss}).\\
Returning to the dynamic model and again exploiting the
renormalization group equation at the fixed point, dimensional
analysis, and the invariant scale transformation we obtain the
universal scaling behaviour of the vertex functions in the asymptotic
limit 
\begin{eqnarray}
  \label{bskalkrit}
  \Gamma_{\tilde{n},n}
  (\{q_{\parallel},{\bf q}_{\perp},\omega\},\tau_{\perp},v_{*},w_{*},\lambda,\rho,\mu)
  & = & \\
  & & \hspace{-200 pt}l^{-\frac{1}{2} \eta (\tilde{n} + n -2)
  -\frac{1}{2} \tilde{n} (d+5) - \frac{1}{2} n(d-3) + d + 5}  \Gamma_{\tilde{n},n}
  \left(\left\{\frac{\textstyle q_{\parallel}}{\textstyle l^{2+\eta}},\frac{\textstyle {\bf q}_{\perp}}{\textstyle l},\frac{\textstyle \omega}{\textstyle l^{4}}\right\},
\frac{\textstyle \tau_{\perp}}{\textstyle l^{2}},v_{*},w_{*},\lambda,\rho,\mu\right)
\; .\nonumber
\end{eqnarray}
Thus, only longitudinal lengths scale anomalously below the upper
critical dimension $ d_{c} = 7 $. From this
equation we derive the scaling form of the density response
function
\begin{equation}
  \label{bsus}
  \chi({\bf q},t) \; = \; f\left(q_{\parallel}^{2} t^{1+\frac{1}{2}
  \eta},q_{\perp}^{2} t^{\frac{1}{2}}\right) 
\end{equation} 
which coincides with the form found for the preceding model. The
positivity of  $ \eta = \frac{1}{5} 
\epsilon + {\cal  O}(\epsilon^{2})  $ signifies superdiffusive behaviour
below $ d_{c} = 7 $. A comparison with the corresponding model without
quenched disorder again shows the quenched disorder to be the reason
for the enhanced spread of fluctuations in the driving force
direction in a dimensional interval that is here, however, $
  5 < d < 7 $.\\
A straightforward extrapolation of the anomaly--exponent into three
dimensions by setting $ \epsilon = 4 $ into the one--loop result gives
the approximate value 
\begin{equation}
    \label{beta31}
    \eta(d=3) \; = \; \frac{4}{5} \; .
  \end{equation}
This numerical value is very distinct from the one--loop and two--loop
values for $ \eta $ in the model with unsymmetric random potential.\\
We now discuss the case that the transverse noise term with $ \gamma
\neq 0 $ of Equation (\ref{udynfunktional}) is generated by coarse
graining even in this symmetric random potential. Then an additional
scaling variable $ \gamma/l^{\phi} $ arises in (\ref{bskalkrit}). The
crossover exponent is here $ \phi = 2 $ and it simply reflects the naive
dimension of the coupling constant $ \gamma $, because $ \gamma $
itself is invariant under the longitudinal scale transformation and
for dimensional reasons the transverse noise term with $ \gamma \neq 0
$ needs no additional renormalization. Since $ \gamma $ is a relevant
variable, for $ \gamma \neq 0 $ the system eventually flows to the
fixed point of the model with unsymmetric random potential.
\subsection{Random Potential With Equally Deep Potential Valleys}
Having no transverse noise according to Equation (\ref{trauschen})
this model is formally obtained from the model with random potential
with equally high mountains by setting the transverse noise
coefficient $ \alpha = 0 $. Then the relevant dynamic functional is
here
\begin{eqnarray}
  \label{tJ} 
  {\cal J}[s,\tilde{s}] & = & \int\!\!
  d^{d}r \left\{\int\!\! dt \left[\tilde{s} \dot{s} + \lambda \tilde{s}
  (\Delta_{\perp}(\Delta_{\perp} - \tau_{\perp}) - \rho
  \Delta_{\parallel}) s + \frac{\textstyle 1}{\textstyle 2} \lambda g
  (\nabla_{\parallel} \tilde{s}) s^{2} \right]\right. \nonumber\\
  & & \hspace{180 pt}\left. - \sigma \left[\lambda \int\!\! dt
  \nabla_{\parallel}   \tilde{s} \right]^{2} \right\} 
 \; .
\end{eqnarray}
The dimensions of lengths, fields, and coupling constants as well as
the upper critical dimension $ d_{c} = 7 $ are as in the
latter model. The symmetry properties of this dynamic functional are
the same as in both preceding models with the exception of the
invariant scale transformation. The lack of the transverse noise term
has the consequence that the dynamic functional is invariant under a
scale transformation depending on two parameters
\begin{eqnarray}
  \label{tskalen}
  r_{\parallel}  \rightarrow  \beta r_{\parallel} & \qquad &
  r_{\perp}  \rightarrow r_{\perp}\\
  \tilde{s}  \rightarrow  \alpha \tilde{s} & \qquad &
  s  \rightarrow  \alpha^{-1}\beta^{-1} s \nonumber\\
  \rho  \rightarrow  \beta^{2} \rho & \qquad & g  \rightarrow 
  \alpha \beta^{2} g  \qquad\qquad \sigma \rightarrow \alpha^{-2}\beta \sigma \; . \nonumber
\end{eqnarray}
Therefore, the combination of coupling constants $ g^{2} \sigma
\rho^{-\frac{5}{2}} $ is the appropriate invariant variable in this model
and is exactly the product of the variables being invariant each in
the latter model, but not here.\\
Because of the quenched disorder we again use the transformation to a
quasi--static Hamiltonian
\begin{equation}
   \label{tquasihamil}
   {\cal H}[\varphi,\tilde{\varphi}] =  \int\!\!
  d^{d}r \left\{\tilde{\varphi}
  (\Delta_{\perp}(\Delta_{\perp} - \tau_{\perp}) - \rho
  \Delta_{\parallel}) \varphi + \frac{1}{2} g
  (\nabla_{\parallel} \tilde{\varphi}) \varphi^{2} + \sigma \tilde{\varphi}
  \Delta_{\parallel} \tilde{\varphi}  \right\} \; .
 \end{equation}
Vertex and Gaussian propagator are the same as in both preceding
models, whereas the Gaussian correlator reads
\begin{equation}
  \label{tpropkorr}
    C_{{\bf q}}   = \frac{2 \sigma q_{\parallel}^{2} }{[{\bf
  q}_{\perp}^{2}({\bf q}_{\perp}^{2}
 + \tau_{\perp}) + \rho  q_{\parallel}^{2}]^{2}}  \; .
\end{equation}
As in the previous model $ \Gamma_{1,1} $, $
\Gamma_{2,0} $, and $ \Gamma_{1,2} $ are primitively divergent. We
have computed them in the lowest nonvanishing order, i.e. we have
performed a one--loop calculation for $\Gamma_{2,0} $ and $
\Gamma_{1,2} $ and a two--loop calculation for $ \Gamma_{1,1} $. The
two--loop calculation involves similar integrals as in
the model with unsymmetric random potential and has also been
performed with the help of inverse Mellin transformation
(cf. Appendix).\\
In dimensional regularization we obtain the primitively divergent
vertex functions, up to higher order terms in $ q $ and $ \epsilon = d_{c} -
d $ 
\begin{eqnarray}
  \label{tvertex}
  \bare{\Gamma}_{1,1}({\bf q}) & = & {\bf
  q}_{\perp}^{2}({\bf q}_{\perp}^{2} + \tau_{\perp}) + \bare{\rho}
  q_{\parallel}^{2}  - 4 \frac{B_{\epsilon}^{2}}{\epsilon^{2}}
  \frac{\bare{g}^{4}\bare{\sigma}^{2}}{\bare{\rho}^{4}} q_{\parallel}^{2}
  \tau_{\perp}^{-\epsilon}( 0 - 0.001799 \epsilon) \\
\bare{\Gamma}_{2,0}({\bf q}) & = & -2 \bare{\sigma}
  q_{\parallel}^{2} -\frac{1}{4} \frac{B_{\epsilon}}{\epsilon}
  \frac{\bare{g}^{2}\bare{\sigma}^{2}}{\bare{\rho}^{\frac{5}{2}}} q_{\parallel}^{2}
  \tau_{\perp}^{-\frac{\epsilon}{2}} \nonumber\\
  \bare{\Gamma}_{1,2}({\bf q}) & = & i \bare{g} q_{\parallel} - i \frac{1}{4}
  \frac{B_{\epsilon}}{\epsilon}
  \frac{\bare{g}^{3}\bare{\sigma}}{\bare{\rho}^{\frac{5}{2}}} q_{\parallel}
  \tau_{\perp}^{-\frac{\epsilon}{2}} \; , \nonumber
\end{eqnarray}
where $ B_{\epsilon} $ is defined by Equation (\ref{bgeofaktor}). The
same redefinition of the coupling constants as in Equation
(\ref{breparameter}) absorbs these divergences and yields in minimal
subtraction 
\begin{eqnarray}
  \label{tZ}
  Z_{\rho} & = & 1 - 0.007196 \frac{v^{2}}{\epsilon} + {\cal O}(v^{3})\\
  Z_{\sigma} & = & 1 - \frac{1}{8} \frac{v}{\epsilon} + {\cal O}(v^{2}) \nonumber\\
  Z_{u} & = & 1 + \frac{1}{4} \frac{v}{\epsilon} + {\cal
  O}(v^{2}) \; ,\nonumber
\end{eqnarray}
where $ v:= B_{\epsilon} u^{2} \sigma \rho^{-\frac{5}{2}} $ is the
dimensionless renormalized variable of the model. This variable is
invariant under the longitudinal scale transformation
(Eq. (\ref{tskalen})) and is the product of $ v $ and $ w $ in the
latter model, but it will here merely be denoted by $ v $ for
simplicity.\\
The renormalization group equation reads
\begin{equation}
  \label{trgg}
  \left [\beta_{v} \frac{\partial}{\partial v} + \rho \zeta_{\rho} \frac{\partial}{\partial
  \rho} +\sigma \zeta_{\sigma} \frac{\partial}{\partial
  \sigma} + \mu \frac{\partial}{\partial \mu}\right ] \Gamma_{\tilde{n},n}
  (\{q_{\parallel},{\bf q}_{\perp}\},\tau_{\perp},v,\rho,\sigma,\mu) = 0
\end{equation}
with the Wilson parameter functions
\begin{eqnarray}
  \label{twilson}
  \beta_{v} & := & \left. \mu \frac{\partial v}{\partial \mu}\right
  |_{o} \; = -v
  \left[\epsilon - \frac{3}{8} v + {\cal O}(v^{2})\right]\\
  \zeta_{\sigma} & := & \left. \mu \frac{\partial \ln \sigma}{\partial
  \mu}\right |_{o} = - \frac{1}{8} v + {\cal O}(v^{2})\nonumber\\ 
  \zeta_{\rho} & := & \left. \mu \frac{\partial \ln \rho}{\partial
  \mu}\right |_{o} = - 0.01439 v^{2} + {\cal O}(v^{3})  \nonumber
\end{eqnarray}
only depending on $ v $.\\
The infrared stable fixed point, being a zero of $\beta_{v} $, is found at
\begin{equation}
  \label{tfixpunkt}
  v_{*} = \frac{\textstyle 8}{\textstyle 3} \epsilon + {\cal O}(\epsilon^{2}) 
  \; .
\end{equation} 
Hence we obtain the anomaly--exponent
\begin{equation}
  \label{teta}
  \eta =  0.0512 \epsilon^{2} + {\cal
  O}(\epsilon^{3}) \; .
\end{equation}
The characteristics of the renormalization group equation are defined
by Equation (\ref{ufluss}) and an additional equation for $
\bar{\sigma}(l) $ that has the same structure as the equation for $
\bar{\rho}(l) $. At the fixed point the solutions for these flow
equations read
\begin{equation}
  \label{tfluss}
  \bar{\rho}(l) = \rho l^{-2 \eta} \qquad\qquad \bar{\sigma}(l) = \sigma
  l^{\zeta_{\sigma *}} \;\; ,
\end{equation} 
where
\begin{equation}
  \label{tsigma}
  \zeta_{\sigma *} \; := \; \zeta_{\sigma}(v_{*}) = - \frac{1}{3}
  \epsilon + {\cal O}(\epsilon^{2}) \;\; . 
\end{equation}
Returning to the dynamic model and combining the renormalization group
equation at the fixed point, dimensional analysis, and the invariant
longitudinal scale transformation we obtain the universal scaling
behaviour of the vertex functions in the asymptotic limit
\begin{eqnarray}
  \label{tskalkrit}
  \Gamma_{\tilde{n},n}
  (\{q_{\parallel},{\bf q}_{\perp},\omega\},\tau_{\perp},v_{*},\lambda,\rho,\sigma,\mu)
  & = & \nonumber\\
  & & \hspace{-80 pt}l^{-\eta (-1 + \frac{3}{2} n -\frac{1}{2}
  \tilde{n}) + \frac{1}{2} \zeta_{\sigma *}(\tilde{n} - n)
  -\frac{1}{2} \tilde{n} (d+5) - \frac{1}{2} n(d-3) + d + 5}\nonumber\\
 & & \hspace{-40 pt} \cdot  \Gamma_{\tilde{n},n}
  \left(\left\{\frac{\textstyle q_{\parallel}}{\textstyle l^{2+\eta}},\frac{\textstyle {\bf q}_{\perp}^{}}{\textstyle l^{}},\frac{\textstyle \omega}{\textstyle l^{4}}\right\},
\frac{\textstyle \tau_{\perp}}{\textstyle l^{2}},v_{*},\lambda,\rho,\sigma,\mu\right)
 \; .
\end{eqnarray}
We again see that only longitudinal lengths scale anomalously. In
contrast to both preceding models there is here an additional exponent
$ \zeta_{\sigma *} $ appearing in the global $ l $--factor. As this
exponent is cancelled for $ \Gamma_{1,1} $, the density response
function nevertheless has the scaling form
\begin{equation}
  \label{tsus}
  \chi({\bf q},t) \; = \; f\left(q_{\parallel}^{2} t^{1+\frac{1}{2}
  \eta},q_{\perp}^{2} t^{\frac{1}{2}}\right) 
\end{equation}
in analogy to both preceding models. As the anomaly--exponent $ \eta = 0.0512
\epsilon^{2} + {\cal  O}(\epsilon^{3}) $ is positive and the upper
critical dimension $ d_{c} = 7 $ is the same as in the preceding
model, all statements concerning superdiffusion and speeding--up of
fluctuations by quenched disorder are here also valid.\\
The straightforward extrapolation of $ \eta $ from $ d_{c} = 7 $ to $
d = 3 $ by setting $ \epsilon = 4 $ produces the approximative value
\begin{equation}
    \label{teta31}
    \eta(d=3) \; = \; 0.82 \; .
  \end{equation}
This numerical value is close to the one--loop value for  $ \eta $ in $
d = 3 $ in the model with random potential with equally high
mountains, but is very far from the one--loop and two--loop values in
the model with unsymmetric random potential.\\
We now investigate the possibility that even in this symmetric random
potential transverse noise terms are produced by coarse
graining. First, we consider the case that transverse noise
proportional to $ \alpha \int d^{d}r[\lambda \int dt \Delta_{\perp}
\tilde{s}]^{2} $ is generated. Then we are in the situation of the
preceding model. If $ \alpha $ is small the system is in
the region of attraction of the degenerate fixed point discussed in
this section. This implies that the transverse noise vanishes under
the renormalization flow and the results of this section remain
valid.\\
Second, if transverse noise proportional to $ \gamma \int
d^{d}r[\lambda \int dt \nabla_{\perp} \tilde{s}]^{2} $  is produced,
an additional 
scaling variable $ \gamma/l^{\phi} $ arises in (\ref{tskalkrit}). Due
to dimensional reasons, this relevant operator needs no additional
renormalization. The invariant
scale transformation according to Equation (\ref{tskalen}) implies $
\gamma \rightarrow \beta^{-1}\alpha^{-2} \gamma $. Thus, the crossover
exponent is related to the exponents $ \eta $ and $ \zeta_{\sigma *} $
by $ \phi = 2 + 2\eta - \zeta_{\sigma *} $. For $ \gamma \neq 0 $ the
system eventually flows to the 
fixed point of the model with unsymmetric random potential, because $
\gamma $ is a relevant variable.

\section{Longitudinal Phase Transition}
The three different realizations of the random potential I--III
described in Section 2 converge to a single model in the region of the
longitudinal phase transition as is sketched in the following.\\
For the three realizations of the random potential the dynamic
functional still containing many irrelevant terms is given by
Eq. (\ref{udynfunktional}) and by Eq. (\ref{bdynfunkt}) with $ \alpha \neq 0
$ and $ \alpha = 0 $, respectively. The longitudinal phase transition
is characterized by $ \tau_{\parallel} \rightarrow 0 $ and finite $
\tau_{\perp} $. The external scale $ \mu^{2} $ here measures small $
\tau_{\parallel} $. Comparing the leading gradient terms in driving
force direction ($ \sim \lambda \tilde{s} \rho \Delta_{\parallel}
\kappa_{\parallel}\Delta_{\parallel} s $) and transverse direction ($
\sim \lambda \tilde{s} \tau_{\perp} \Delta_{\perp} s $) we find for
length scales 
\begin{equation}
  \label{lskalierung}
  r_{\parallel} \sim \mu^{-1} \qquad r_{\perp} \sim \mu^{-2} \; .
\end{equation}
As the dynamic functional is dimensionless 
the coupling constants of the transverse noise scale as $ \gamma \sim \mu^{-2}
$, $ \alpha \sim \mu^{-6} $, i.e., only the longitudinal noise is
relevant. Hence, only a single model is required to describe  the
longitudinal phase 
transition, independent of the kind of the random potential. Further,
the dimensions of fields and the other coupling constants are 
\begin{eqnarray}
  \label{ldimension}
  \lambda t \sim \mu^{-4}\qquad &\qquad s \sim \mu^{d -
 \frac{7}{2}}\qquad &\qquad
 \tilde{s} \sim
  \mu^{d + \frac{5}{2}} \nonumber\\
  \kappa_{\perp} \sim \mu^{-4}\qquad & \qquad\kappa \sim \mu^{-2}\qquad & \qquad\sigma
  \sim \mu^{0} \nonumber\\
  \qquad &\qquad  g \sim \mu^{\frac{13}{2} - d}\; ,\qquad & \qquad
\end{eqnarray}
indicating that $ \kappa_{\perp} $ and $ \kappa $ are irrelevant and
that 
\begin{equation}
  \label{lgrenzdim}
  d_{c} = 6.5 
\end{equation}
is the upper critical dimension of this model.\\
After a suitable scale change of fields and lengths the relevant
dynamic functional for the longitudinal phase transition is
\begin{eqnarray}
  \label{lJ}
  {\cal J}[s,\tilde{s}] & = & \int\!\!
  d^{d}r \left\{\int\!\! dt \left[\tilde{s} \dot{s} + \lambda \tilde{s}
  (- \Delta_{\perp} + \rho
  \Delta_{\parallel}(\rho
  \Delta_{\parallel} - \tau_{\parallel})) s + \frac{\displaystyle 1}{\displaystyle 2} \lambda g
  (\nabla_{\parallel} \tilde{s}) s^{2} \right]\right. \nonumber\\
  & & \hspace{180 pt}\left. - \sigma \left[\lambda \int\!\! dt
  \nabla_{\parallel}   \tilde{s} \right]^{2} \right\} 
 \; .
\end{eqnarray}
The symmetry properties are the same as in the last model of Section 3
inclusively the invariant longitudinal scale transformation, because
transverse noise exists in neither model. As a consequence $ g^{2}\sigma
\rho^{-\frac{5}{2}} $ is the appropriate invariant variable of this
model.\\
Due to the quenched disorder, we transform to a quasistatic
Hamiltonian
\begin{equation}
   \label{lquasihamil}
   {\cal H}[\varphi,\tilde{\varphi}] =  \int\!\!
  d^{d}r \left\{\tilde{\varphi}
  (-\Delta_{\perp} + \rho
  \Delta_{\parallel}(\rho \Delta_{\parallel} - \tau_{\parallel})) \varphi + \frac{1}{2} g
  (\nabla_{\parallel} \tilde{\varphi}) \varphi^{2} + \sigma \tilde{\varphi}
  \Delta_{\parallel} \tilde{\varphi}  \right\} \; .
 \end{equation}
From this equation we directly read off the elements of perturbation
theory, i.e., the vertex $ -igq_{\parallel} $ and the Gaussian
propagator and correlator
\begin{eqnarray}
  \label{lpropkorr}
   G_{{\bf q}} & = & \frac{1}{{\bf
  q}_{\perp}^{2} + \rho  q_{\parallel}^{2}(\rho  q_{\parallel}^{2} +\tau_{\parallel}) } \\ 
  C_{{\bf q}}  & = &\frac{2 \sigma q_{\parallel}^{2} }{[{\bf
  q}_{\perp}^{2} + \rho  q_{\parallel}^{2}(\rho  q_{\parallel}^{2} +\tau_{\parallel})]^{2}}  \; .\nonumber
\end{eqnarray} 
The vertex functions $\Gamma_{1,1} $, $ \Gamma_{2,0} $, and $
\Gamma_{1,2} $ are primitively divergent. The singular parts of
$\Gamma_{1,1} $ are proportional to $
q_{\parallel}^{4} $ and $ q_{\parallel}^{2} $, the singular parts of 
$ \Gamma_{2,0} $ and $ \Gamma_{1,2} $ are proportional to $
q_{\parallel}^{2} $ and $ q_{\parallel} $, respectively.\\
In a one--loop calculation we obtain the primitively divergent
regularized vertex functions in bare quantities up to higher orders in
$ q $ and $ \epsilon = d_{c} - d $
\begin{eqnarray}
  \label{lvertex}
  \bare{\Gamma}_{1,1}({\bf q}) & = & {\bf
  q}_{\perp}^{2} + \bare{\rho}
  q_{\parallel}^{2}(\bare{\rho}
  q_{\parallel}^{2} + \bare{\tau}_{\parallel}) + \frac{C_{\epsilon}}{\epsilon}
  \frac{\bare{g}^{2}\bare{\sigma}}{\bare{\rho}^{\frac{3}{2}}} q_{\parallel}^{2}
  \bare{\tau}_{\parallel}^{-\epsilon}(\frac{1}{6}\bare{\rho} q_{\parallel}^{2}
  + 0 \cdot \bare{\tau}_{\parallel}) \\
\bare{\Gamma}_{2,0}({\bf q}) & = &  -2 \bare{\sigma}
  q_{\parallel}^{2} -\frac{1}{6} \frac{C_{\epsilon}}{\epsilon}
  \frac{\bare{g}^{2}\bare{\sigma}^{2}}{\bare{\rho}^{\frac{5}{2}}} q_{\parallel}^{2}
  \bare{\tau}_{\parallel}^{-\epsilon}\nonumber\\
  \bare{\Gamma}_{1,2}({\bf q}) & = & i \bare{g} q_{\parallel} - i \frac{1}{6}
  \frac{C_{\epsilon}}{\epsilon}
  \frac{\bare{g}^{3}\bare{\sigma}}{\bare{\rho}^{\frac{5}{2}}} q_{\parallel}
  \bare{\tau}_{\parallel}^{-\epsilon} \; , \nonumber
\end{eqnarray}
where
\begin{equation}
  \label{lgeofaktor}
  C_{\epsilon} = \frac{1}{(2\pi)^{d}} {\cal O}_{d-1} \Gamma(1 +
  \epsilon) \Gamma(\frac{11 - 2\epsilon}{4}) \Gamma(\frac{5 - 2\epsilon}{4})
\end{equation}
is a suitably chosen $ \epsilon $--dependent factor. These divergences
are absorbed in the redefinition of the coupling constants
\begin{equation}
  \label{lreparameter}
  \bare{\rho} \; = \; Z_{\rho} \rho \qquad \bare{\tau}_{\parallel} \; = \;
  Z_{\tau_{\parallel}} \tau_{\parallel} \qquad \bare{\sigma} \; = \;
  Z_{\sigma} \sigma \qquad \bare{g}  \;  =  \;
  \mu^{\epsilon} Z_{u} u \; .
\end{equation}
In minimal subtraction we obtain the $ Z $--factors as a function of
the dimensionless renormalized invariant variable $ v:= C_{\epsilon}
u^{2} \sigma \rho^{-\frac{5}{2}} $
\begin{eqnarray}
  \label{lZ}
  Z_{\rho} & = & 1  - \frac{1}{12} \frac{v}{\epsilon} + {\cal O}(v^{2}) \\
  Z_{\tau_{\parallel}} & = &  1  + \frac{1}{12} \frac{v}{\epsilon} + {\cal O}(v^{2}) \nonumber\\
  Z_{\sigma} & = & 1 - \frac{1}{12} \frac{v}{\epsilon} + {\cal O}(v^{2}) \nonumber\\
  Z_{u} & = & 1 + \frac{1}{6} \frac{v}{\epsilon} + {\cal
  O}(v^{2}) \; .\nonumber
\end{eqnarray}
Here, the renormalization group equation is of the form
\begin{equation}
  \label{lrgg}
  \left [\beta_{v} \frac{\partial}{\partial v} + \rho \zeta_{\rho} \frac{\partial}{\partial
  \rho} +\sigma \zeta_{\sigma} \frac{\partial}{\partial
  \sigma} + \kappa \tau_{\parallel}\frac{\partial}{\partial \tau_{\parallel} } + \mu \frac{\partial}{\partial \mu}\right ] \Gamma_{\tilde{n},n}
  (\{q_{\parallel},{\bf q}_{\perp}\},\tau_{\parallel},v,\rho,\sigma,\mu) = 0 \; 
\end{equation}
with the Wilson functions
\begin{eqnarray}
  \label{lwilson}
  \beta_{v} & := & \left. \mu \frac{\partial v}{\partial \mu}\right
  |_{o} \; = -2v
  \left[\epsilon - \frac{11}{24} v + {\cal O}(v^{2})\right]\\
  \zeta_{\sigma} & := & \left. \mu \frac{\partial \ln \sigma}{\partial
  \mu}\right |_{o} = - \frac{1}{6} v + {\cal O}(v^{2})\nonumber\\ 
  \zeta_{\rho} & := & \left. \mu \frac{\partial \ln \rho}{\partial
  \mu}\right |_{o} = - \frac{1}{6} v + {\cal O}(v^{2})  \nonumber\\
  \kappa & := & \left. \mu \frac{\partial \ln \tau_{\parallel}}{\partial
  \mu}\right |_{o}  = \frac{1}{6} v + {\cal O}(v^{2}) \; . \nonumber
\end{eqnarray}
In addition to the flow equations for $ \bar{v}(l) $, $
\bar{\rho}(l) $, $ \bar{\sigma}(l) $, and $ \bar{\mu}(l) $ being of
the same form as in the preceding model there is a flow equation for $
\bar{\tau}_{\parallel}(l) $ that reads
\begin{equation}
  \label{lfluss}
  \frac{d}{dl} \ln \bar{\tau}_{\parallel}(l) = \kappa(\bar{v}(l)) \; .
\end{equation}
In the scaling limit $ l \ll 1 $ that corresponds to the relations
\begin{equation}
  \label{lskalenlimes}
  \left |\frac{q_{\parallel}}{\mu} \right | \ll 1 \qquad
  \left |\frac{{\bf q}_{\perp}}{\mu^{2}} \right | \ll 1 \qquad
  \left |\frac{\tau_{\parallel}}{\mu^{2}} \right | \ll 1
\end{equation}
$ \bar{v}(l) $ flows to an infrared stable fixed point $ v_{*} $ which
is directly obtained as a zero of $ \beta_{v} $
\begin{equation}
  \label{lfixpunkt}
  v_{*} = \frac{\textstyle 24}{\textstyle 11} \epsilon + {\cal O}(\epsilon^{2})
  \; .
\end{equation}
At the fixed point the solutions of the flow equations are given by
Equation (\ref{tfluss}) and by $ \bar{\tau}_{\parallel}(l) =
\tau_{\parallel} l^{\kappa_{*}} $ with $ \kappa_{*} := \kappa(v_{*})
$.
Inserting the fixed point value for $ v_{*} $ into $ \zeta_{\rho} $ we 
find the anomaly--exponent
\begin{equation}
  \label{leta}
  \eta = \frac{\textstyle 2}{\textstyle 11} \epsilon + {\cal
  O}(\epsilon^{2}) \; . 
\end{equation}
The fixed point values of the other parameter functions are 
\begin{equation}
  \label{lparfix}
  \zeta_{\sigma *} = - \frac{4}{11} \epsilon + {\cal O}(\epsilon^{2})
  \qquad\qquad \kappa_{*} = \frac{4}{11} \epsilon + {\cal O}(\epsilon^{2})
  \; .
\end{equation}
Returning to the dynamic model and combining the solutions of the
renormalization group equation at the fixed point with a dimensional
analysis and the invariant scale transformation we finally arrive at
the universal scaling behaviour of the vertex functions at the
longitudinal phase transition
\begin{eqnarray}
  \label{lskalkrit} 
  \Gamma_{\tilde{n},n}
  (\{q_{\parallel},{\bf q}_{\perp},\omega\},\tau_{\parallel},v_{*},\lambda,\rho,\sigma,\mu)
  & = & \nonumber\\
  & & \hspace{-80 pt}l^{-\eta (-1 + \frac{3}{2} n -\frac{1}{2}
  \tilde{n}) + \frac{1}{2} \zeta_{\sigma *}(\tilde{n} - n)
  -\frac{1}{2} \tilde{n} (2d+5) - \frac{1}{2} n(2d-7) + 2d + 3}\nonumber\\
 & & \hspace{-70 pt} \cdot  \Gamma_{\tilde{n},n}
  \left(\left\{\frac{\textstyle q_{\parallel}}{\textstyle l^{1+\eta}},
\frac{\textstyle {\bf q}_{\perp}^{}}{\textstyle l^{2}},\frac{\textstyle \omega}{\textstyle l^{4}}\right\},
\frac{\textstyle \tau_{\parallel}}{\textstyle l^{2-\kappa_{*}}},v_{*},\lambda,\rho,\sigma,\mu\right)
 \; .
\end{eqnarray}
In addition to longitudinal lengths the critical parameter $
\tau_{\parallel} $ also exhibits anomalous scaling behaviour. In
analogy to the models in Section 3 we derive the scaling form of the
density response function
\begin{equation}
  \label{lchi}
\chi({\bf q},t) \; = \; f\left(q_{\parallel}^{2} t^{\frac{1}{2}(1 +
  \eta)},q_{\perp}^{2} t \right) \; .  
\end{equation}
While the system is normally diffusive with respect to the transverse
directions, in the critical longitudinal direction typical length
squares scale for long times as
\begin{equation}
  \label{llangskal}
  < r_{\parallel}^{2} > \; \sim \; t^{\frac{1}{2}(1 + \eta)} \; ,
\end{equation}
i.e. in comparison to the naively critical $ t^{\frac{1}{2}} $
(corresponding to the naive dynamical critical exponent $ z = 4 $ from
model B \cite{HH}) the spread of fluctuations in the driving force
direction is enhanced, due to the positive $ \eta = \frac{2}{11}
\epsilon + {\cal O}(\epsilon^{2}) $.\\
A comparison with the corresponding model for the longitudinal phase
transition in a system without quenched disorder \cite{JS2} shows the
surprising result that the longitudinal phase transition is here
continuous, evidenced by the existence of an infrared stable fixed point,
whereas in the model without quenched disorder there is no infrared
stable fixed point.

\section{Summary and Outlook}
\begin{figure}[!ht]
\begin{center}
\input{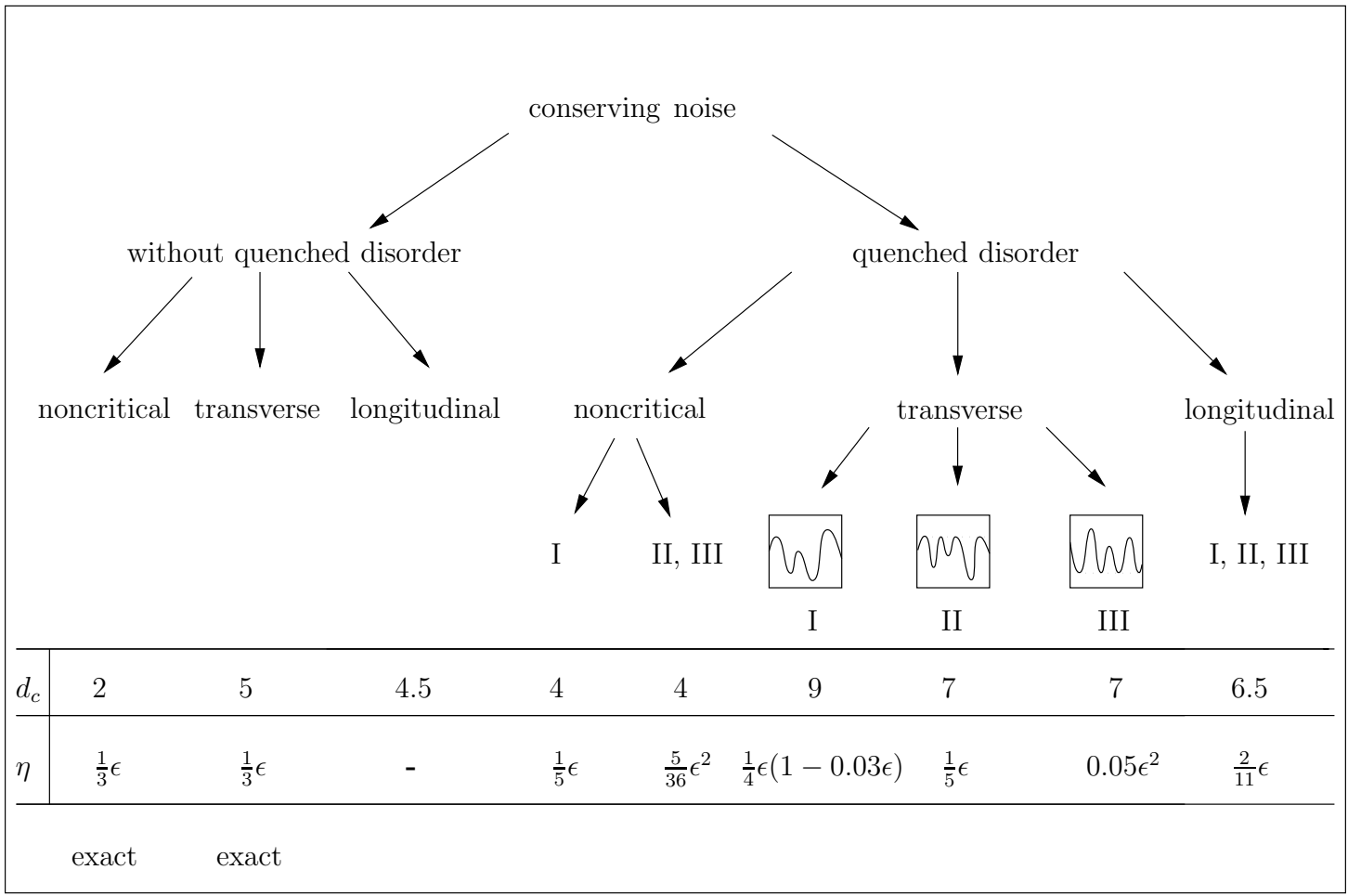}
\end{center}
\caption{Model class of driven diffusive systems with conserving noise.}
      \label{ahnen1} 
\end{figure}

We have analyzed the transverse and longitudinal phase transition in
uniformly driven diffusive systems with quenched disorder. These
systems show a wide variety of 
possible scenarios, because the symmetry properties of the random
potential are an additional distinguishing feature to which
universality class a model belongs. In the region of the transverse phase
transition the three different random potentials I--III (Fig. \ref{randompot})
actually define three different models and in the noncritical region
they still define two different models all of which are part of different
universality classes.\\
Together
with earlier investigations the renormalization group studies of an
entire model class are hereby completed. This model class includes the
driven diffusive systems with and without quenched disorder both in
the critical regions of the transverse and longitudinal phase
transition and in the noncritical region. Fig. \ref{ahnen1} gives a
graphical overview of the model class, where the single models are
ordered chronologically from left to the right. The models in an
ordered substrate \cite{JS1,JS2,LC}, i.e. without quenched disorder, had
been studied prior to this work for all three regions of the phase
diagram mentioned above. The models for a system with quenched
disorder in the noncritical region \cite{BJ1} had also been
investigated. The other models of Fig. \ref{ahnen1} have been studied in
the present paper.\\
The upper critical dimension $ d_{c} $ is different from model to model
and varies from 2 to 9. Below $ d_{c} $, the vertex functions being
typical statistical quantities of such systems show universal
anomalous scaling behaviour on large length and time scales. The
deviation from pure diffusive behaviour is characterized by the
anomaly--exponent $ \eta $. It indicates how strongly longitudinal
lengths scale anomalously. In Fig. \ref{ahnen1} the result for $ \eta $ in
the highest calculated loop order is given for every model. Note that,
due to Galilean invariance, $ \eta $ is even exact in all models
without quenched disorder.\\
We emphasize the following results:\\
In all models of this model class (except for the longitudinal phase
transition in a system without quenched disorder) the
anomaly--exponent $ \eta $ is positive implying superdiffusive
spreading of density fluctuations in the driving force direction.\\
A model with quenched disorder always has a higher upper critical
dimension than the corresponding model without quenched disorder. Due to
the field theoretic results, in the dimension interval between these
two upper critical dimensions the quenched disorder is the reason for
superdiffusive spread of density fluctuations in the high temperature
region and at the transverse phase transition, respectively. Notice
that the anomalous diffusion of 
fluctuations is not directly connected with the behaviour of transport
coefficients relating the mean current to the mean density, because a
fluctuation--dissipation theorem (Einstein relation) does not hold in
this strong nonequilibrium situation with quenched disorder. Within
this model class we have not found a qualitative argument why disorder
generates superdiffusion. We mention, however, a study of a
one dimensional driven lattice gas 
\cite{KFer,E} where quenched disorder is not, as here,
spatially fixed, but associated
with moving particles. The authors found a superdiffusive spread of a jam
behind the slowest particle and a superdiffusive spread of free
spacing in front of it. Whether this situation can be transferred to
our models by identifying the deepest potential valley and the highest
potential mountain, respectively, with the
slowest particle, should be a topic of further investigations,
especially Monte Carlo simulations.\\
For the longitudinal phase transition there is no uniform
statement. In the model without quenched disorder \cite{JS2} no
infrared stable fixed point has been found and some analytic arguments
point to a discontinuous phase transition, whereas an according two
dimensional lattice gas demonstrates a continuous phase transition in
Monte Carlo simulations \cite{BassZ}. In the model with quenched disorder
studied here, however, we find an infrared stable fixed point and thus
a continuous longitudinal phase transition.\\
Despite the partially high upper critical dimensions it appears
that the anomaly--exponent (when existing) may be extrapolated quite
accurately into low dimensions. Two observations lend support to this
procedure: first, the two--loop
correction to $ \eta $ in the critical transverse model from Section
3.1  is small and second, the coefficients of $ \eta $ are small in all
models.\\
For the two--loop calculation in critical models with quenched
disorder a new technique has been developed enabling us to manipulate
mixed $ {\bf q}^{4} $--propagators and --correlators. This technique
is based on an inverse Mellin transformation and is
described for one model at the transverse phase transition in the
Appendix. This method is applicable to all critical models with
quenched disorder of this model class, but has only been performed for
two models (Sec. 3.1 and 3.3). Furthermore we expect it to be useful
for two--loop calculations in other physical problems where $ {\bf
  q}^{4} $--propagators are involved.\\
Monte Carlo simulations of two--dimensional driven diffusive lattice
gases without quenched disorder \cite{Leung1,Leung2,KLS1,KLS2} are in
excellent agreement with field 
theoretic predictions for the transverse phase transition and the
noncritical region \cite{JS1,JS2,LC}. For the transverse phase
transition in systems 
with quenched disorder, however, there is a simulation study \cite{LF}
that is, for two reasons, hardly compatible with the models
investigated here by field theory. First,
the quenched disorder was there modelled by randomly blocked sites and
not by random potential barriers between the sites. Second, the
concentration of blocked sites is so small that only the crossover
behaviour between a system with and without quenched disorder was
observed. Further, we mention a recent simulation for a noncritical
system with quenched disorder \cite{BarT}.\\
It is desirable to compare the field theoretic results obtained here
for the various critical models with corresponding Monte Carlo
simulations that are 
still to be done. These simulations are a nontrivial challenge,
because besides the average over a huge number of realizations of the
quenched disorder it is to pay attention to the fact that the periodic
boundary conditions usually used let pass a particle repeatedly
through the system and let it see the same quenched disorder as
before. By this unwanted correlations of the randomness enter into the
simulational results that make it difficult to compare them with the
field theoretic results that are based on the assumption of
uncorrelated disorder. Therefore Monte Carlo simulations of driven
diffusive systems with open boundaries already done for systems
without quenched disorder \cite{JOe2} seem to be more appropriate.\\
We finally remark that we have extended the model class investigated
here in the way that we allow for random particle sources and drains
in the diffusive systems. The
noncritical model without quenched disorder with such a particle
nonconserving randomness had already been studied
\cite{HK,BJ2}. Moreover we
have analyzed the influence of nonconserving noise onto the
critical behaviour in systems with and without
quenched disorder which is demonstrated in a paper soon to be
published \cite{BJ4}.

\appendix
\section{Two--Loop Calculation of $ \Gamma_{1,1} $}
\begin{figure}[!ht]
  \begin{center}
    \leavevmode
\input{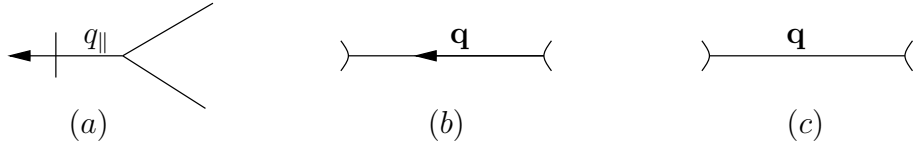}    
    \caption{The graphical elements of perturbation theory in the
      quasi--static model: (a)
      vertex, (b) Gaussian propagator, and (c) Gaussian correlator.}
    \label{graphelem}
  \end{center}
\end{figure}

\begin{figure}[!ht]
  \begin{center}
    \leavevmode
\begin{tabular}{ccc}
\includegraphics*[scale = 1]{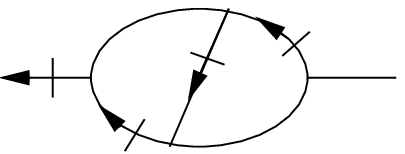} & \includegraphics*[scale =
1]{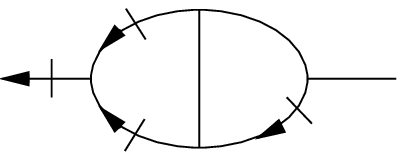} & \includegraphics*[scale = 1]{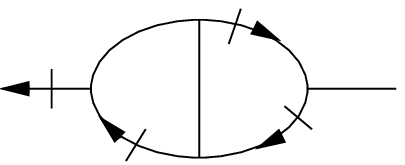}\\ 
$ B^{(1,1)}_{1} $ &  $ B^{(1,1)}_{2} $ & $ B^{(1,1)}_{3} $\\[12 pt]
\includegraphics*[scale = 1]{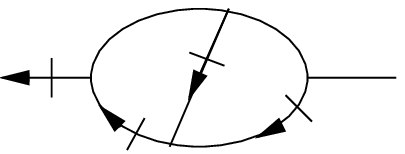} & \includegraphics*[scale =
1]{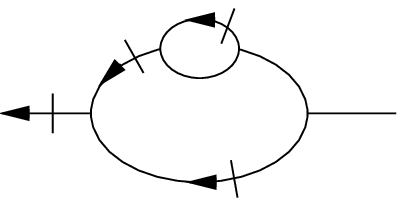} & \includegraphics*[scale = 1]{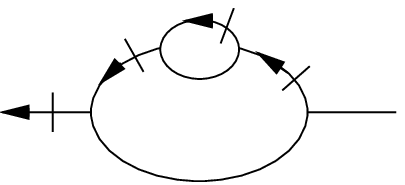}\\ 
$ B^{(1,1)}_{4} $ &  $ B^{(1,1)}_{5} $ & $ B^{(1,1)}_{6} $\\[12 pt]
\includegraphics*[scale = 1]{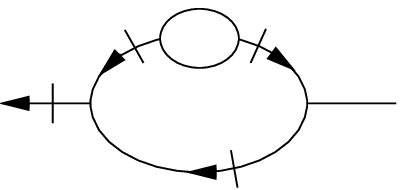} & \includegraphics*[scale =
1]{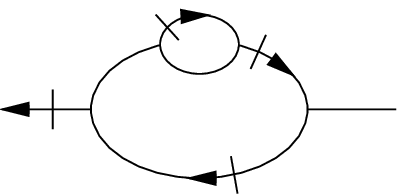} &  \\ 
$ B^{(1,1)}_{7} $ &  $ B^{(1,1)}_{8} $ & 
\end{tabular} 
    \caption{All two--loop diagrams of $ \Gamma_{1,1} $ obeying
      causality. The symmetry factor of $ B^{(1,1)}_{7} $ is $
      \frac{1}{2} $, whereas it is 1 for all other diagrams.}
    \label{2loop11}
  \end{center}
\end{figure}

\begin{figure}[!ht]
  \begin{center}
    \leavevmode
\input{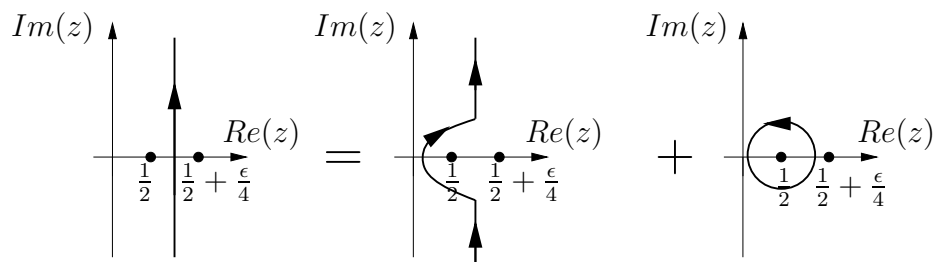} 
    \caption{The complex integration path lying between two poles is
      decomposed into two parts.}
    \label{intpath}
  \end{center}
\end{figure}

In the Appendix we explicitly show the two--loop calculation for the
model that describes the transverse phase transition in an unsymmetric
random potential (cf. Section 3.1). The graphical elements of the
diagrammatic perturbation expansion are shown in Fig. \ref{graphelem}
directly for the quasi--static model which is here used to
facilitate the calculation. The $ \tilde{\varphi} $--legs are
indicated by an arrow and the $ q_{\parallel} $ of the vertex by a
dash perpendicular to the propagator line. The mathematical
expressions for the graphical elements are given by Equation
(\ref{quasipropkorr}).\\
In this model we have to calculate the primitively divergent vertex
functions $ \Gamma_{1,1} $ and $ \Gamma_{1,2} $. There are eight
two--loop diagrams contributing to $ \Gamma_{1,1} $ (Fig. \ref{2loop11}) and
25 two--loop diagrams contributing to $ \Gamma_{1,2} $ 
each of which obeys causality that forbids closed propagator
loops. Momentum conservation demands that at each vertex the sum over
all wave vectors is zero. Evaluation of these diagrams requires
integration over all internal wave vectors.\\ 
While the one--loop calculation is easy to perform by standard
methods, the two--loop calculation involving mixed $ {\bf q}^{4}
$--propagators requires more sophisticated tools and has not been
described in literature. We introduce a technique that is based on an
inverse Mellin transformation.\\
For $ \Gamma_{1,1} $ we have to compute the diagrams $ B^{(1,1)}_{1} $ to $
B^{(1,1)}_{8} $ from Fig. \ref{2loop11}. We denote the external momentum by
$ {\bf q} $ and the internal ones to be integrated over by $ {\bf p} $ and $
{\bf k} $. As $ \Gamma_{1,1} $ is quadratically divergent because of
CP--symmetry and  
dimensional reasons and the external $
\tilde{\varphi} $ already provides a factor $ q_{\parallel} $, the
parts of the integrands being proportional to $ q_{\parallel} $
contain all singularities. Therefore, the integrands are first
expanded with respect to the external momentum $ q_{\parallel} $ to
first order. For simplicity, we now substitute $ \rho^{\frac{1}{2}} p_{\parallel} \rightarrow
p_{\parallel}, \; \rho^{\frac{1}{2}} k_{\parallel} \rightarrow
k_{\parallel} $ and from now on we omit the superscript ``$ \; \bare{} $ "
to characterize unrenormalized quantities.\\
In the next step all even powers of $ p_{\parallel} $ and $
k_{\parallel} $ in the numerator of the integrands are written as $
p_{\parallel}^{2} = [{\bf p}_{\perp}^{2}({\bf p}_{\perp}^{2} +
  \tau_{\perp}) + p_{\parallel}^{2}] - {\bf p}_{\perp}^{2}({\bf p}_{\perp}^{2} +
  \tau_{\perp}) $ ($ k_{\parallel}^{2} $ analogously). The first term
  of the right hand side cancels against factors in the
  denominator. For all odd 
  powers the half of the integrand is reflected with respect to $
k_{\parallel} $. This step reduces the superficial degree of
divergence of the $ {\bf p} $-- and $
{\bf k} $--integration by 2. It is allowed because the
integration runs over the whole $ k_{\parallel} $--axis.\\
Then all integrals are reduced to the two types of integrals 
\begin{eqnarray}
  \label{I}
  I(\alpha,\beta,\gamma,\delta,\mu,\nu)& := & \\
 & &\hspace{-70 pt} \int_{\bf p} \frac{{\bf p}_{\perp}^{2\alpha} }
  {[{\bf p}_{\perp}^{2}({\bf p}_{\perp}^{2} +
  \tau_{\perp}) + p_{\parallel}^{2}]^{\beta}} 
 \int_{\bf k} \frac{({\bf p}_{\perp} -{\bf k}_{\perp})^{2 \gamma}}{[({\bf p}_{\perp}
  -{\bf k}_{\perp})^{4} + (p_{\parallel} - k_{\parallel})^{2}]^{\delta}}
\frac{{\bf k}_{\perp}^{2\mu} }{[{\bf k}_{\perp}^{4} +
 k_{\parallel}^{2}]^{\nu}} \nonumber
\end{eqnarray}\\[-20 pt]
\begin{eqnarray*}
F(A;\alpha,\beta,\gamma,\delta ,\mu,\nu)& := & \int_{\bf p}
  \frac{p_{\parallel}^{A} {\bf p}_{\perp}^{2\alpha} }{[{\bf p}_{\perp}^{2}({\bf p}_{\perp}^{2} +
  \tau_{\perp}) + p_{\parallel}^{2}]^{\beta}} \nonumber\\
 & & \hspace{-110 pt}\cdot 
\int_{\bf k} \frac{k_{\parallel}{\bf k}_{\perp}^{2\mu} }{[{\bf k}_{\perp}^{4} +
  k_{\parallel}^{2}]^{\nu}} 
 \left[\frac{({\bf p}_{\perp}-{\bf k}_{\perp})^{2
  \gamma}}{[({\bf p}_{\perp}-{\bf k}_{\perp})^{4}+(p_{\parallel}
 -k_{\parallel})^{2}]^{\delta}} - \frac{({\bf p}_{\perp}-{\bf k}_{\perp})^{2
  \gamma}}{[({\bf p}_{\perp}-{\bf k}_{\perp})^{4}+(p_{\parallel}
 +k_{\parallel})^{2}]^{\delta}}
  \right]\nonumber
\end{eqnarray*}
the arguments of which can be 0, 1, 2,...\\
As the integrands are already expanded with respect to $ q_{\parallel}
$, the whole integration over $ {\bf p} $ and $ {\bf k} $ is at most
logarithmically divergent. The integration variables are chosen such
that the $ {\bf p} $--subintegration is always primitively
convergent, i.e., the naive dimension $ \delta_{{\bf p}} $ of this
integration (measured in powers of the external momentum scale $ \mu
$) is negative. In the integrals of both types the $
{\bf k} $--integration is at most logarithmically divergent.\\
Due to these naive dimensions of the $ {\bf p} $--, $ {\bf k} $-- and the
whole integration it is possible (and necessary for the following
calculus) to set $ \tau_{\perp} = 0 $ in the $ {\bf k} $--integration,
because the $ \tau_{\perp} \neq 0 $--parts only provide convergent
contributions.\\
The sum of the two--loop diagrams for $ \Gamma_{1,1} $ expressed by
the integral types $ I $ and $ F $ reads
\begin{eqnarray}
  \label{a2}
  \sum_{i=1}^{8} B^{(1,1)}_{i} & = & 4\frac{g^{4}}{\rho^{2}} q_{\parallel}^{2} [3
  I(1,3,0,1,1,2) - 3 I(1,3,0,1,3,3) - 2 I(3,4,0,1,1,2) \nonumber\\[4 pt]
  &+ & \hspace{-3mm} 2
  I(3,4,0,1,3,3) - 2 \tau_{\perp} I(2,4,0,1,1,2) + 2 \tau_{\perp}
  I(2,4,0,1,3,3) \nonumber\\
  &+ & \hspace{-3mm} \frac{1}{2} I(0,2,1,2,1,2) - \frac{3}{2}
  I(2,3,1,2,1,2) + I(4,4,1,2,1,2) \nonumber\\
  & - &\hspace{-3mm}  F(3;1,4,0,1,1,3) - \frac{1}{2}
  F(1;1,3,0,1,1,3) + F(1;1,4,1,2,0,1)]  .\nonumber\\
\end{eqnarray}
By factorizing the $ {\bf q}^{4} $--denominators in $ I $ und $ F $
into transverse and longitudinal parts we are in the position to apply
the successful methods that are used for $ {\bf q}^{2} $--propagators
and --correlators. This factorization is done by the Mellin
transformation. We demonstrate the method for the integral type $ I $
in detail. The calculation of $ F $ goes analogously.\\
The Mellin transformation of the function $ (a + x)^{-\alpha} $ is 
\begin{equation}
  \label{mellin}
   \int_{0}^{\infty} \! dx x^{t-1} (a + x)^{-\alpha} =
  \frac{\Gamma(t)\Gamma(\alpha - t)}{\Gamma(\alpha)} a^{t - \alpha} \;
  ,
\end{equation}
where the conditions $ a > 0 $ and  $ 0 < Re(t) < Re(\alpha) $ must
be fulfilled \cite{Erdelyi}. The corresponding inverse Mellin
transformation
\begin{equation}
  \label{invmellin}
  (a + x)^{-\alpha} = \int_{t_{0} - i\infty}^{t_{0} + i\infty}
\frac{dt}{2\pi i} \left(\frac{\Gamma(t)\Gamma(\alpha - t)}{\Gamma(\alpha)}
a^{t - \alpha}\right) x^{-t} \; ,
\end{equation}
where the integration path parallel to the imaginary axis is
restricted by $ 0 < t_{0} < Re(\alpha) $ \cite{Erdelyi}, proves to be the
appropriate tool to factorize the denominators in $ I $.\\
First we only treat the $ {\bf k} $--integration of the integral type
$ I $ (\ref{I}) and apply the inverse Mellin transformation to both
denominators 
\begin{eqnarray}
  \label{ik}
  I_{{\bf k}} & := & \int_{\bf k}  \frac{({\bf p}_{\perp} -{\bf k}_{\perp})^{2 \gamma}}{[({\bf p}_{\perp}
  -{\bf k}_{\perp})^{4} + (p_{\parallel} - k_{\parallel})^{2}]^{\delta}}
\frac{{\bf k}_{\perp}^{2\mu} }{[{\bf k}_{\perp}^{4} + k_{\parallel}^{2}]^{\nu}}  \\
& = & \int_{t_{0} - i\infty}^{t_{0} + i\infty}
\frac{dt}{2\pi i} \int_{s_{0} - i\infty}^{s_{0} + i\infty}
\frac{ds}{2\pi i} \frac{\Gamma(t) \Gamma(\delta - t) \Gamma(s)
  \Gamma(\nu - s)}{\Gamma(\delta)\Gamma(\nu)}\nonumber\\
& & \hspace{8 mm} \cdot \int_{\bf k} \frac{1}{[k_{\parallel}^{2}]^{s} [(p_{\parallel} -
  k_{\parallel})^{2}]^{t}}
 \frac{1}{[{\bf k}_{\perp}^{2}]^{2(\nu - s) - \mu}
  [({\bf p}_{\perp} - {\bf k}_{\perp})^{2}]^{2(\delta - t) - \gamma}}
  \; .\nonumber
\end{eqnarray}
In the integrand transverse and longitudinal momenta are now separated
and only quadratic. The $ {\bf k} $--integration can now be performed
with the help of the usual Feynman relations
\begin{eqnarray}
  \label{feyn}
 \frac{1}{A^{\alpha}} & = & \frac{1}{\Gamma(\alpha)} \int_{0}^{\infty} ds
  s^{\alpha - 1} e^{-s A} \\
\frac{1}{\prod_{i}^{} A_{i}^{\alpha_{i}}}
& = & \frac{\Gamma(\sum_{i}^{}\alpha_{i})}
{\prod_{i}^{}\Gamma(\alpha_{i})}
\int_{0}^{1} \prod_{i}^{} dx_{i} x_{i}^{\alpha_{i} - 1} \frac{\delta(
\sum_{i}^{}x_{i} - 1)}{[\sum_{i}^{} x_{i}
  A_{i}]^{\sum_{i}^{}\alpha_{i}}} \; ,\nonumber
\end{eqnarray}
where  $ \Gamma(\alpha) $ is Euler's $ \Gamma $ function. The result
reads
\begin{eqnarray}
  \label{ik2}
  I_{{\bf k}} & = & \frac{1}{2} \frac{{\cal O}_{d-1}}{(2\pi)^{d}}
  \Gamma(\frac{1}{2}) \Gamma(\frac{d - 1}{2}) \int_{t_{0} - i\infty}^{t_{0} + i\infty}
\frac{dt}{2\pi i} \int_{s_{0} - i\infty}^{s_{0} + i\infty}
\frac{ds}{2\pi i} \frac{\Gamma(t) \Gamma(\delta - t) \Gamma(s)
  \Gamma(\nu - s)}{\Gamma(\delta)\Gamma(\nu)}\nonumber\\
& & \cdot \frac{\Gamma(s + t - \frac{1}{2})\Gamma(\frac{1}{2} -
  s)\Gamma(\frac{1}{2} - t)}{\Gamma(s)\Gamma(t)\Gamma(1 - s - t)} \frac{\Gamma(2\nu +
    2\delta - \mu - \gamma - \frac{d-1}{2} - 2(s + t))}{\Gamma(2(\nu -
  s)-\mu)\Gamma(2(\delta - t)-\gamma)}\nonumber\\
& & \cdot \frac{\Gamma(\frac{d-1}{2}-2(\nu - s)+\mu)
  \Gamma(\frac{d-1}{2}-2(\delta -
  t)+\gamma)}{\Gamma(d-1-2\nu-2\delta+\mu+\gamma+2(s + t))}\nonumber\\
& & \cdot |p_{\parallel}|^{1 - 2(s+t)} \; |{\bf p}_{\perp}|^{d - 1 + 4(s + t -
  \nu - \delta) + 2(\gamma + \mu)} \; .
\end{eqnarray}
The $ {\bf k} $--integral exists under the conditions 
\begin{eqnarray}
\label{ik3}
2 (s_{0} +  t_{0}) & > & 1\\
s_{0} < \frac{1}{2} \qquad\qquad t_{0} & < & \frac{1}{2}\nonumber\\
4(\nu + \delta - (s_{0} + t_{0})) - 2(\mu + \gamma) & > & d - 1\nonumber\\
d - 1 - 4(\nu - s_{0}) + 2\mu & > & 0\nonumber\\
d - 1 - 4(\delta - t_{0}) + 2\gamma & > & 0 \nonumber
\end{eqnarray}
which prevent UV and IR divergences, respectively, at the $
k_{\parallel} $-- and $ 
{\bf k}_{\perp} $--integration. These conditions restrict the complex
integration 
paths of the $ s $-- and $ t $--integration which are here dependent
on the spatial dimension, due to the dimensional regularization.\\
The result of the $ {\bf k} $--integration is according to Equation
(\ref{ik2}) proportional to powers of $ |p_{\parallel}| $ and $ |{\bf
  p}_{\perp}| $ with exponents that are dependent on $ s $, $ t $, and
$ d $. Thus, the remaining $ {\bf p} $--integration is analogous to
the one--loop problem, up to changed exponents. It is
straightforwardly performed and we obtain for $ I $
\begin{eqnarray}
\label{ik4}
I(\alpha,\beta,\gamma,\delta,\mu,\nu) & = & C_{d}
  \frac{\Gamma(2(\beta +
    \delta + \nu) - (\alpha + \gamma + \mu) - (d + 1))}{\tau_{\perp}^{2(\beta +
    \delta + \nu) - (\alpha + \gamma + \mu) - (d + 1)}}\nonumber\\
& & \hspace{-35 mm} \cdot \int_{t_{0} - i\infty}^{t_{0} + i\infty} \!\!\!
\frac{dt}{2\pi i} \int_{s_{0} - i\infty}^{s_{0} + i\infty} \!\!\!
\frac{ds}{2\pi i} \Gamma(2\nu +
    2\delta - \mu - \gamma - \frac{d-1}{2} - 2(s+t))\Gamma(s + t - \frac{1}{2})\nonumber\\
& & \hspace{-35 mm} \cdot \Gamma(\delta -
  t) \Gamma(\nu - s) \Gamma(\frac{1}{2} -
  s) \Gamma(\frac{1}{2} -
  t) \frac{\Gamma(d+s+t+\alpha+\gamma+\mu-\beta-2\delta-2\nu)}
  {\Gamma(d-1-2\nu-2\delta+\mu+\gamma+2(s+t))}\nonumber\\
& &   \cdot \frac{\Gamma(\frac{d-1}{2}-2(\nu-s)+\mu)
  \Gamma(\frac{d-1}{2}-2(\delta-t)+\gamma)}
  {\Gamma(2(\nu-s)-\mu)\Gamma(2(\delta-t)-\gamma)} \; ,
\end{eqnarray}
where $ C_{d} = \frac{1}{4} \left(\frac{{\cal
      O}_{d-1}}{(2\pi)^{d}}\right)^{2}\frac{\Gamma(\frac{1}{2})
 \Gamma(\frac{d - 1}{2})}{\Gamma(\beta)\Gamma(\delta)\Gamma(\nu)} $ is
      a constant that is different for every $ I $ and depends on the
      dimension. The conditions for the existence of the $ {\bf p}
      $--integration are
\begin{eqnarray}
  \label{ik5}
  1 - (s_{0} + t_{0}) & > & 0 \\
  d + s_{0} + t_{0} + \alpha + \gamma + \mu - \beta - 2\delta - 2\nu &
  > & 0 \nonumber\\
  2(\beta + \delta + \nu) - (\alpha + \gamma + \mu) - (d + 1) & > & 0
  \; . \nonumber
\end{eqnarray}
After the momentum integrations there remain parameter integrations
with respect to the Mellin variables $ s $ and $ t $ over parallels to
the imaginary axis. The complex integration paths and the spatial
dimension $ d $ are restricted by the conditions (\ref{ik3}) and
(\ref{ik5}). With $ \epsilon = 9 - d $ we obtain from the first four
inequalities of (\ref{ik3})
\begin{equation}
  \label{ik6}
  \frac{1}{2} < s_{0} + t_{0} < \delta + \nu - \frac{1}{2} (\gamma +
  \mu) - 2 + \frac{\epsilon}{4}  \; ,
\end{equation}
where $ s_{0} < \frac{1}{2},\; t_{0} < \frac{1}{2} $. All other
conditions are satisfied for all $ I $ of Equation (\ref{a2}), if $ 0
< \epsilon < 2 $.\\
Since the sum $ s_{0} + t_{0} $ appears in the inequality (\ref{ik6}) we
transform to the new integration variables 
\begin{equation}
  \label{ik7}
  z := s + t \qquad w := \frac{1}{2} (s - t) \; .
\end{equation}
The integration paths are now determined by 
\begin{eqnarray}
  \label{ik8}
  \frac{1}{2} \quad < \quad z_{0} & < & \delta + \nu - \frac{1}{2} (\gamma +
  \mu) - 2 + \frac{\epsilon}{4}\\
   |w_{0}| & < & \frac{1}{4} \; .\nonumber 
\end{eqnarray}
After this substitution Equation (\ref{ik4}) gives 
\begin{eqnarray}
  \label{ik9}
I(\alpha,\beta,\gamma,\delta,\mu,\nu) & = & C_{d} \frac{\Gamma(2(\beta +
    \delta + \nu) - (\alpha + \gamma + \mu) - (d + 1))}{\tau_{\perp}^{2(\beta +
    \delta + \nu) - (\alpha + \gamma + \mu) - (d + 1)}} \\
& &\hspace{-38 mm} \cdot \int_{z_{0} - i\infty}^{z_{0} + i\infty} \frac{dz}
{2\pi i} \int_{w_{0} - i\infty}^{w_{0} + i\infty} 
\frac{dw}
{2\pi i} \Gamma(2\nu +
    2\delta - \mu - \gamma - \frac{d-1}{2} - 2z)\Gamma(z-\frac{1}{2})
    f(z,w;\epsilon)  ,\nonumber
\end{eqnarray}
where the abbreviation 
\begin{eqnarray}
        \label{ik10}
       f(z,w;\epsilon) & := & \Gamma(\delta -
  \frac{z}{2} + w) \Gamma(\nu - \frac{z}{2} - w) \Gamma(\frac{1}{2} -
  \frac{z}{2} - w)\Gamma(\frac{1}{2} -
  \frac{z}{2} + w) \nonumber\\
& & \hspace{-1 mm}\cdot\frac{\Gamma(\frac{8-\epsilon}{2}-2(\nu-\frac{z}{2}-w)+\mu)
  \Gamma(\frac{8-\epsilon}{2}-2(\delta-\frac{z}{2}+w)+\gamma)}
  {\Gamma(2(\nu-\frac{z}{2}-w)-\mu)\Gamma(2(\delta-\frac{z}{2}+w)-\gamma)}\nonumber\\
& & \hspace{-1 mm}\cdot\frac{\Gamma(9-\epsilon
+z+\alpha+\gamma+\mu-\beta-2\delta-2\nu)} 
  {\Gamma(8-\epsilon
-2\nu-2\delta+\mu+\gamma+2z)}
      \end{eqnarray}
denotes the part of the integrand that is free of poles.\\
To extract the divergent parts of $ I $ we have to distinguish two
cases.\\
Case 1: The $ {\bf k} $--integration is primitively convergent.\\
In this case the parameters satisfy
\begin{equation}
  \label{ik11}
  \nu + \delta - \frac{1}{2}(\gamma + \mu) = 3 \; ,
\end{equation}
which is true for $I(0,2,1,2,1,2), \; I(2,3,1,2,1,2) $, and $
I(4,4,1,2,1,2) $ from Equation (\ref{a2}). According to Equation
(\ref{ik8}) this means for the constant $ z_{0} $ which determines the
integration path
\begin{equation}
   \label{ik12}
   \frac{1}{2} < z_{0} < 1 + \frac{\epsilon}{4} \; .
 \end{equation}
For these $I $ the whole integration over $ {\bf p} $ und $
{\bf k} $ is logarithmically divergent so that their parameters
fulfill the equation
\begin{equation}
  \label{ik15}
  2(\beta +
    \delta + \nu) - (\alpha + \gamma + \mu) - (d + 1) = \epsilon \; .
\end{equation}
Thus, the coefficient $ \Gamma(\epsilon) = \frac{1}{\epsilon}
\Gamma(1+\epsilon) $ of the double integral over $ z $ und $ w
$ (\ref{ik9}) contains an $
\epsilon $--pole, whereas the double integral itself is convergent, as
the integrand is free of poles even in the limit $ \epsilon
\rightarrow 0 $ and the real part of the integration path can be
chosen between $\frac{1}{2} $ and 1 according to Equation
(\ref{ik12}). The double integral in Equation (\ref{ik9}) can
therefore be computed at $ \epsilon = 0 $, because together with the $
\epsilon $--pole as coefficient only convergent contributions are produced for
$ \epsilon \neq 0 $. The convergence of the double integral is ensured
by the asymptotic behaviour of the $ \Gamma $ function with complex
arguments \cite{Gradshteyn}
\begin{equation}
  \label{gammaasymp}
 \lim_{|y| \rightarrow \infty} |\Gamma(x + iy)| \;
  e^{\frac{\pi}{2}|y|} \; |y|^{\frac{1}{2} - x} \; = \;
  (2\pi)^{\frac{1}{2}} \; .
\end{equation}
Due to the accumulation of $ \Gamma $ functions we are not able to
perform the integrations with respect to  $ z $ and  $ w $
analytically. Hence, each of the three $ I $ is computed
numerically for $ \epsilon = 0 $.\\ 
Case 2: The $ {\bf k} $--integration is logarithmically divergent.\\
In this case the parameters are restricted to
\begin{equation}
  \label{ik17}
  \nu + \delta - \frac{1}{2}(\gamma + \mu) = \frac{5}{2} \; ,
\end{equation}
which is correct for the remaining $I $ from Equation (\ref{a2}). According to Equation
(\ref{ik8}) the integration path is here restricted by
\begin{equation}
   \label{ik18}
   \frac{1}{2} < z_{0} < \frac{1}{2} + \frac{\epsilon}{4} \; .
 \end{equation}
In the limit $ \epsilon \rightarrow 0 $ the integration path is trapped by
this condition between the poles of the
integrand  (\ref{ik9}) at $ z = \frac{1}{2} $ and $ z =
\frac{1}{2}  + \frac{\epsilon}{4} $. In order to extract the $
\epsilon $--poles the complex integration path is decomposed into two
parts (as shown graphically in Fig. \ref{intpath}): The first part is
a line parallel to the imaginary axis with an arc to the left of the
singularity $ z = \frac{1}{2} $, while the second part is a circle
around $ z = \frac{1}{2} $. The $ z $--integral over the circle gives
the residuum at $ z = \frac{1}{2} $.\\
This way we obtain from Equation (\ref{ik9})
\begin{eqnarray}
  \label{ik19}
I(\alpha,\beta,\gamma,\delta,\mu,\nu) & = & C_{d} \frac{\Gamma(2(\beta +
    \delta + \nu) - (\alpha + \gamma + \mu) - (d + 1))}{\tau_{\perp}^{2(\beta +
    \delta + \nu) - (\alpha + \gamma + \mu) - (d + 1)}} \\
& &\hspace{-35 mm} \cdot \left[\int_{z_{0}^{'} - i\infty}^{z_{0}^{'} + i\infty} \frac{dz}
{2\pi i} \int_{w_{0} - i\infty}^{w_{0} + i\infty} 
\frac{dw}
{2\pi i} \Gamma(1 + \frac{\epsilon}{2} - 2z)\Gamma(z-\frac{1}{2})
    f(z,w;\epsilon)\right.\nonumber\\
 & &\hspace{25 mm}\left. + \; \Gamma(\frac{\epsilon}{2})
 \int_{w_{0} - i\infty}^{w_{0} + i\infty} \frac{dw}
{2\pi i}f(z=\frac{1}{2},w;\epsilon)\right]  ,\nonumber
\end{eqnarray}
Naturally, the integration path of the $ z $--integration is again
chosen as a parallel to the imaginary axis, whose real part $
z_{0}^{'} $ now lies between $ -\frac{1}{2} $ and $ \frac{1}{2}
$. After the integration path is changed, both the double integral
over $ z $ and $ w $ and the single integral over $ w $ are eventually
convergent. Only the coefficients of the integrals contain the $
\epsilon $--poles. For the further calculation we have again to
distinguish two cases.\\
a) The whole integration is logarithmically divergent.\\
This is the case for the integrals $ I(1,3,0,1,1,2),\; I(1,3,0,1,3,3),\;
I(3,4,0,1,1,2)$, and $ I(3,4,0,1,3,3) $ from Equation(\ref{a2}). Their
parameters satisfy Equation (\ref{ik15}) so that in front of  the entire
bracket there is an $ \epsilon $--pole due to $ \Gamma(\epsilon) $.\\
For every single $ I $ the double integral is finally calculated
for $ \epsilon = 0 $ numerically and provides the coefficients of simple $ \epsilon
$--poles. Parts of the integrand with $ \epsilon \neq 0 $ lead to
convergent contributions and are therefore omitted.\\
The coefficient $ \Gamma(\epsilon)\Gamma(\frac{\epsilon}{2}) $ of the
simple integral, however, contains an $ \epsilon^{2} $--pole. Hence,
the integrand $ f(z=\frac{1}{2},w;\epsilon) $ must be expanded with
respect to $ \epsilon $ up to the first order
\begin{eqnarray}
        \label{ik20}
       f(z=\frac{1}{2},w;\epsilon) & = & \frac{\Gamma(\frac{19}{2}-\epsilon
+\alpha+\gamma+\mu-\beta-2\delta-2\nu)}
  {\Gamma(9-\epsilon
-2\nu-2\delta+\mu+\gamma)}\Gamma(\delta -
  \frac{1}{4} + w)\nonumber\\
& & \hspace{-35 mm} \cdot \Gamma(\nu - \frac{1}{4} - w) \Gamma(\frac{1}{4} - w)\Gamma(\frac{1}{4} + w) 
 \frac{\Gamma(\frac{9-\epsilon}{2}-2(\nu-w)+\mu)
  \Gamma(\frac{9-\epsilon}{2}-2(\delta+w)+\gamma)}
  {\Gamma(2(\nu-\frac{1}{4}-w)-\mu)\Gamma(2(\delta-\frac{1}{4}+w)-\gamma)}\nonumber\\
&  &\hspace{-35 mm} = \frac{\Gamma(\beta - \frac{1}{2} - \epsilon)}{\Gamma(4 -
  \epsilon)} \Gamma(\delta -
  \frac{z}{2} + w) \Gamma(\nu - \frac{z}{2} - w) \Gamma(\frac{1}{2} -
  \frac{z}{2} - w)\Gamma(\frac{1}{2} -
  \frac{z}{2} + w)\nonumber\\
& & \hspace{-25 mm} \cdot \left[1 - \frac{\epsilon}{2}(\Psi(2\delta-\frac{1}{2}+2w-\gamma) +
  \Psi(2\nu-\frac{1}{2}-2w-\mu))+
  {\cal O}(\epsilon^{2})\right] \; ,
      \end{eqnarray}
where $ \Psi(x) = \frac{\Gamma^{'}(x)}{\Gamma(x)} $ denotes Euler's $
\Psi $ function and the relations (\ref{ik15}) and (\ref{ik17}) have
been used. The integral of the zeroth order in $ \epsilon $ provides
the coefficients of the $ \epsilon^{2} $--poles and can even be
executed analytically \cite{Gradshteyn}
\begin{eqnarray}
  \label{ik21}
  \int_{w_{0} - i\infty}^{w_{0} + i\infty} \frac{dw}
{2\pi i} \Gamma(\delta -
  \frac{1}{4} + w) \Gamma(\nu - \frac{1}{4} - w) \Gamma(\frac{1}{4} -
  w)\Gamma(\frac{1}{4} + w) & = & \\
 & & \hspace{-40 mm} \frac{\Gamma(\delta)\Gamma(\delta+\nu-\frac{1}{2})\Gamma(\nu)\Gamma(\frac{1}{2})}{\Gamma(\delta+\nu)} \; .\nonumber
\end{eqnarray}
The integral of the first order in $ \epsilon $ yields
the coefficients of the $ \epsilon $--poles and is numerically
calculated for every single $ I $ that belongs to this case 2 a).\\
b) The whole integration is primitively convergent.\\
This statement is true for $ \tau_{\perp} I(2,4,0,1,1,2) $ and $\tau_{\perp}
  I(2,4,0,1,3,3) $ in Equation (\ref{a2}). Here the parameters have
  the property
\begin{equation}
  \label{ik23}
  2(\beta +
    \delta + \nu) - (\alpha + \gamma + \mu) - (d + 1) = 1 + \epsilon
    \; ,
\end{equation}
i.e. in front of the brackets in Equation (\ref{ik19}) the coefficient
is $ \Gamma(1 + \epsilon) $ and consequently there is no $ \epsilon $--pole.\\
Therefore the double integral needs not to be computed, as it only leads to
convergent contributions. The simple integral can be evaluated
at $ \epsilon = 0 $ because the coeffient contains only a simple $
\epsilon $--pole. Due to the relation (\ref{ik17}) that is also valid here,
the simple integral is reduced to the one already solved in Equation
(\ref{ik21}) analytically.\\
Now all integrals of type $ I $ from Equation (\ref{a2}) have been
calculated. The way to solve the integrals of type $ F $ is step by step
analogous to the method presented for type $ I $. $ F(3;1,4,0,1,1,3) $
and $ F(1;1,3,0,1,1,3) $ from Equation (\ref{a2}) belong to case 1,
whereas $ F(1;1,4,1,2,0,1) $ belongs to case 2  a).\\
Finally we present the sum of the two--loop diagrams for $ \Gamma_{1,1}
$  up to convergent parts (cf. (\ref{uvertex}))
\begin{equation}
  \label{ik35}
  \sum_{i=1}^{8} B^{(1,1)}_{i} = 4\frac{g^{4}}{\rho^{2}}
  q_{\parallel}^{2} A_{\epsilon}^{2}
  \frac{\tau_{\perp}^{-\epsilon}}{\epsilon^{2}} \left(\frac{1}{18} -
0.01199 \epsilon\right) \; .
\end{equation}
The 25 two--loop diagrams of the primitively divergent $ \Gamma_{1,2}
$ again only lead to integrals of type $ I $ and $ F $
\begin{eqnarray}
  \label{ik36}
  \sum_{i=1}^{25} B^{(1,2)}_{i} & = & -4i\frac{g^{5}}{\rho^{3}} q_{\parallel} [3
  I(1,3,0,1,1,2) -  I(1,3,1,2,2,2) - 3 I(3,4,0,1,1,2) \nonumber\\
  &+ & \hspace{-3mm} 
  I(3,4,1,2,2,2) - 2 I(1,3,0,1,3,3) + 2 I(3,4,0,1,3,3)\nonumber\\
 & - &3 \tau_{\perp} I(2,4,0,1,1,2) + 2 \tau_{\perp}
  I(2,4,0,1,3,3) + \tau_{\perp} I(2,4,1,2,2,2)\nonumber\\
  &- & \hspace{-3mm} \frac{1}{2} I(0,2,1,2,1,2) +
  I(2,3,1,2,1,2) -\frac{1}{2} I(4,4,1,2,1,2) \nonumber\\
  & - &\hspace{-3mm}  F(3;1,4,0,1,1,3) +
  F(3;0,3,1,2,1,3) + 2 F(3;1,5,1,2,0,1)]  ,\nonumber\\
\end{eqnarray}
 which can be
evaluated with the methods demonstrated for $ \Gamma_{1,1} $. This
concludes the two--loop calculation.

{\bf Acknowledgement}\\
The authors thank K. Oerding and B. Schmittmann for illuminating
discussions. This work has been supported by the DFG under SFB 237
(Unordnung und Grosse Fluktuationen). 
\bibliographystyle{unsrt}     
\bibliography{lit}           
\newpage
\newpage
\newpage
\newpage
\newpage
\newpage
\end{document}